%% file: main.tex
\documentclass[nature,longbibliography,superscriptaddress, twocolumn]{revtex4-2}  
\input{preambel}

\begin{document}

    \makeatletter
    \def\@caption#1[#2]#3{%
      \par\addcontentsline{\csname ext@#1\endcsname}{#1}{%
      \protect\numberline{\csname the#1\endcsname}{#2}}%
      \begingroup
        \@parboxrestore
        \leftskip=0pt
        \rightskip=\@flushglue
        \parindent=0pt
        \parfillskip=0pt
        \noindent
        {\bfseries \csname fnum@#1\endcsname:} #3\par
      \endgroup}
    \makeatother

    \input{title}
    \input{abstract_new2}
    \maketitle

    \section*{Introduction}
    \label{sec:intro}

    \begin{figure*}[t!]
            \centering
            \includegraphics[width=.99\textwidth]{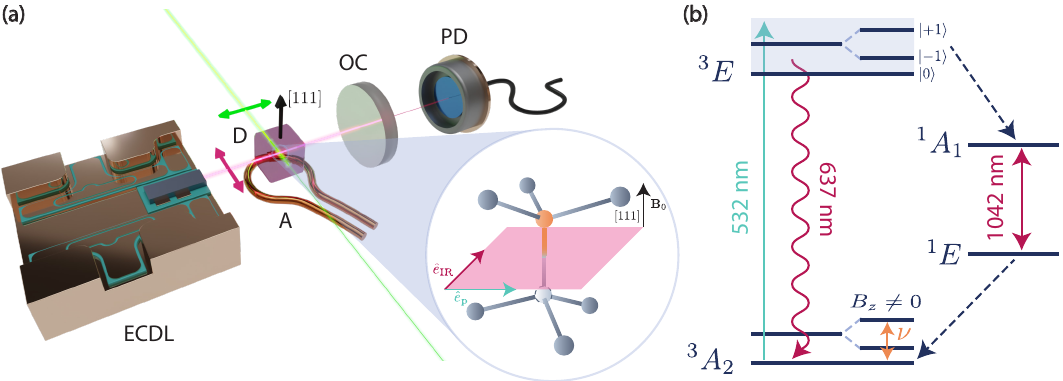}
            \caption[3D model and level scheme]{(a) Schematic of the LICAM setup. An external-cavity diode laser (ECDL) provides optical gain at $1042\,$nm. A collimating lens (not shown) directs the edge emission through the diamond (D), which is optically pumped at $532\,$nm. The diamond is cut along the $[111]$ crystal axis (black arrow), which enables perfect cross-sectioning of the linearly polarized pump and probe beams with the $[111]$ NV$^-$ resonances. A coupling mirror (OC) closes the cavity, and the optical output power is detected by a photodiode (PD). Magnetic resonances are driven by MW radiation from the antenna (A). (b) Simplified NV$^-$ energy level scheme. The ground-state triplet manifold $^3\mathrm{A}_2$ is off-resonantly excited to the excited-state manifold $^3\mathrm{E}$, which fluoresces in red while predominantly preserving the spin state. An external magnetic field lifts the degeneracy of the Zeeman sublevels $\ket{\pm1}_\mathrm{m_S}$, allowing selective excitation by resonant MW fields of frequency $\nu$. From the excited state, these sublevels preferentially decay via nonradiative intersystem crossing into the singlet state $^1\mathrm{A}_1$, involving an optical transition at $1042\,$nm. Probing this transition enables optical detection of the magnetic resonances.}
            \label{fig:1}
        \end{figure*}

        Optical quantum sensors, including optically pumped magnetometers (OPMs) and nitrogen-vacancy (NV$^-$) centers in diamond, represent a promising platform for biomedical sensing applications, due to their ability to detect magnetic fields under ambient conditions \cite{Budker.2007,Aslam.2023}. OPMs rely on atomic vapors and achieve ultrahigh magnetic sensitivities at the attotesla level \cite{Budker.2007, Dang.2010} approaching those of superconducting quantum interference devices \cite{Dumas.2021} and enabling wearable, yet bulky, and mobile platforms for magnetoencephalography \cite{Boto.2018}, magnetocardiography \cite{EscalonaVargas.2025} and magnetomyography \cite{Nordenstrom.2024}. Further reduction of sensor footprints leads to reduced sensitivities as a result of unavoidable lower active sensing volumes and therefore requires compensation schemes to maintain high sensitivities \cite{Mitchell.2020}. A promising approach to overcome this limitation is the use of optical cavities to enhance light-matter interaction, which has been demonstrated, yet with insufficient sensitivity enhancements \cite{Crepaz.2015}.
        
        An alternative sensing platform are NV$^-$-based sensors, for which photonic integration is more advanced and which enable highly sensitive measurements of magnetic fields, electric fields, and temperature over a broad frequency range from DC to GHz \cite{Taylor.2008,Dolde.2011,Kucsko.2013,Jakobi.2017}. Diamond as a host material offers exceptional biocompatibility making NV$^-$ sensors highly attractive for diverse research fields and applications in life sciences, material sciences and more \cite{Glenn.2018, Casola.2018, Qiu.2022}. NV$^-$-based quantum sensors have demonstrated key milestones in biomedical sensing, including the detection of a single neuron's action potential \cite{Barry.2016} and millimeter-scale magnetocardiography \cite{Arai.2022}.
        In parallel, major advances have been made toward on-chip integration, fiber packaging, and scalable architectures \cite{Ibrahim.2021, Sturner.2021, Bopp.2025, Gazzano.2017}. 
        
        Despite these achievements, significant sensitivity enhancements are still required to make use of integrated NV$^-$-sensors for demanding biomedical applications such as magnetoencephalography. Similar to OPMs, a viable approach is the use of an optical cavity to enhance sensitivities. Cavity-based implementations have been successfully demonstrated for NV$^-$ ensembles \cite{Jensen.2014,Chatzidrosos.2017}. Nevertheless, these methods have so far not reached fundamental sensitivity limits.

        One of the most sensitive spectroscopic techniques for enhancing weak absorption signals is intracavity absorption spectroscopy, in which the spectrum of a laser with a broadband  gain profile is highly responsive to frequency-dependent losses in the resonator  \cite{Pakhomycheva.2021,Kimble.1980, Fjodorow.2024}. Sensitivity enhancements up to seven orders of magnitude have been demonstrated \cite{Baev.1999}.

        In this work, we combine the principles of intracavity absorption spectroscopy and laser threshold magnetometry \cite{Jeske.2016}, demonstrating laser intracavity absorption magnetometry (LICAM) to improve the sensitivity of transmission-based optical quantum sensors using a self-sustained infrared external-cavity diode laser (ECDL). 
        Laser threshold magnetometry was proposed to improve the sensitivity of NV$^-$-based magnetometers. It exploits a magnetic-field-dependent optical \textit{gain} provided by the NV$^-$ triplet transition inside a laser resonator. When operated close to the lasing threshold, this additional gain can push the laser above threshold, potentially enabling near-unity spin contrast and high magnetic sensitivities \cite{Jeske.2016}. Numerical studies have extended this concept to absorption-based schemes mediated by infrared absorption of the NV$^-$ singlet transition \cite{Dumeige.2019}, as well as off-resonant pump absorption of the triplet transition \cite{RamanNair.2020, Webb.2021}. More recently, schemes at the laser threshold have been demonstrated experimentally for both stimulated emission of the visible transition \cite{Hahl.2022, Lindner.2024} and infrared absorption \cite{Gottesman.2024, schall.2026}. These implementations have relied on external-cavity surface-emitting diode lasers, which require optical seeding with high pump powers, thereby limiting scalability and integration. 
        
        Whereas laser threshold magnetometry primarily exploits magnetic-field-dependent threshold shifts or threshold switching, LICAM uses the MW-induced NV\(^{-}\) singlet absorption as a continuous intracavity loss modulation and enables a compact, self-contained, and power-efficient magnetometer design with prospects for on-chip integration. We benchmark LICAM against an equivalent single-pass configuration and demonstrate up to a $475$-fold enhancement of the optically detected spin contrast and a $180$-fold improvement in magnetic sensitivity. Importantly, general enhancements are not only achieved near the lasing threshold but also well above it. We model LICAM using single-mode diode laser rate equations and find excellent agreement with our experimentally determined enhancement factors. Using realistic parameters in our model we determine operational requirements for reaching magnetic sensitivities on the femtotesla level.

        \section*{Results} 
        \label{sec:results}

        \begin{figure}[ht!]	
        \centering
        \includegraphics[width=0.49\textwidth]{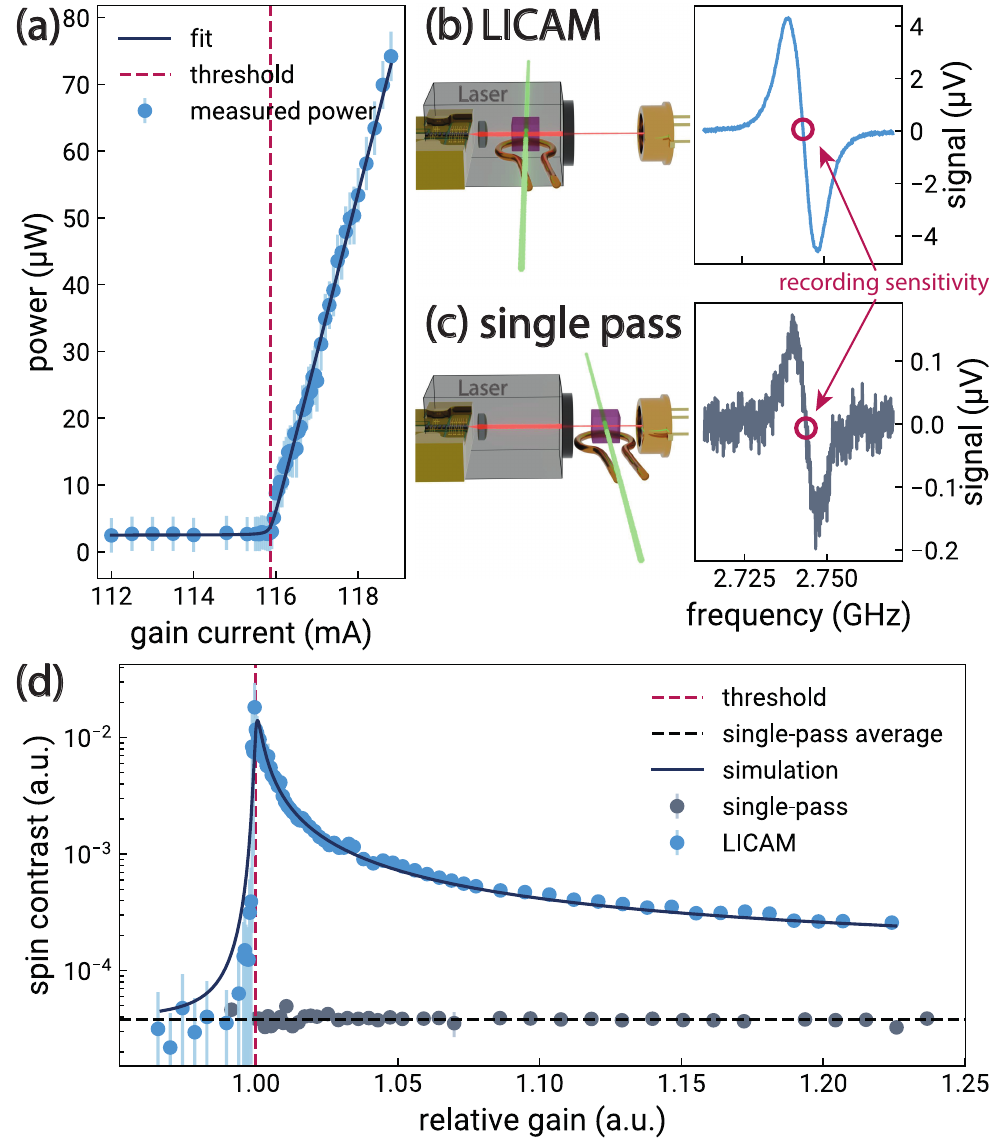}
        \caption[Characteristic laser curve, ODMR resonance and ODMR contrast]{(a) Characteristic laser output curve of the LICAM sensor, showing a threshold at $116\,$mA (red dashed line). (b) LICAM configuration and ODMR trace. (c) Single-pass configuration and ODMR trace. ODMR trace in (b) and (c) of the $([111],-1_{\mathrm{m}_\mathrm{S}})$-resonance recorded at a relative gain current of $1.03$. (d) ODMR contrast as a function of normalized gain current. The gain is normalized to the respective lasing threshold. The solid dark-blue line in (a) represents a rate-equation fit (Eq.(\ref{eq:Pout})), while the solid line in (d) is the corresponding contrast prediction based on Eq.(\ref{eq:depth}). The dashed black line in (d) indicates the mean contrast value for the single-pass configuration.} 
        \label{fig:2}
        \end{figure}

        The LICAM sensor consists of an electrically driven ECDL incorporating a diamond sample that is part of the optical feedback path in the cavity (see FIG. \ref{fig:1}). The laser emission wavelength is selected by a distributed Bragg reflector (DBR) grating in the chip, which can be temperature-tuned to match the NV$^-$ singlet absorption line at $1042\,$nm. The edge emission of the chip is collimated by an intracavity lens, and the diamond is pumped perpendicularly with a $532\,$nm laser. A microwave (MW) field drives the NV$^-$ spin transitions, and sinusoidal frequency modulation of the MW signal enables lock-in detection of optically detected magnetic resonances (ODMRs) through the optical output of the cavity. Varying the ECDL gain current (FIG. \ref{fig:2}(a)) allows to study spin contrast and magnetic sensitivity around and above the lasing threshold (for details see \hyperref[sec:mm]{Materials and Methods}).
        
        \subsection*{Enhancement of spin contrast}
        FIG. \ref{fig:2}(b) shows the spin contrast as a function of gain current. The spin contrast sharply increases from about $4\times10^{-5}$ below the lasing threshold to $1.8\times10^{-2}$ at the threshold. Increasing the gain current above the threshold current leads to a rapid contrast decrease saturating at about $3\times10^{-4}$. To benchmark the LICAM performance, we perform conventional single-pass infrared absorption measurements under identical conditions, with the diamond placed outside the cavity. In this configuration, the contrast remains constant at a mean value of $3.8\times10^{-5}$, independent of the gain current. This behaviour is expected, as the spin contrast does not depend on the probe power \cite{Bopp.2025}. We note that the LICAM contrast below threshold approaches the single-pass contrast, confirming that measurement parameters including the active sensing volume are equivalent in both configurations. Comparing to the single-pass contrast, the LICAM contrast reveals a $475$-fold enhancement at the lasing threshold. Moreover, the measured enhancement as a function of gain current agrees excellently with simulations based on single-mode diode-laser rate equations (see section \hyperref[sec:rateq]{simulation}). Notably, LICAM yields contrast enhancements not only at the lasing threshold but also well above it, maintaining roughly an order-of-magnitude improvement.

        \subsection*{Enhancement of magnetic sensitivity}
        
        \begin{figure}[ht!]	
        \centering
        \includegraphics[width=0.49\textwidth]{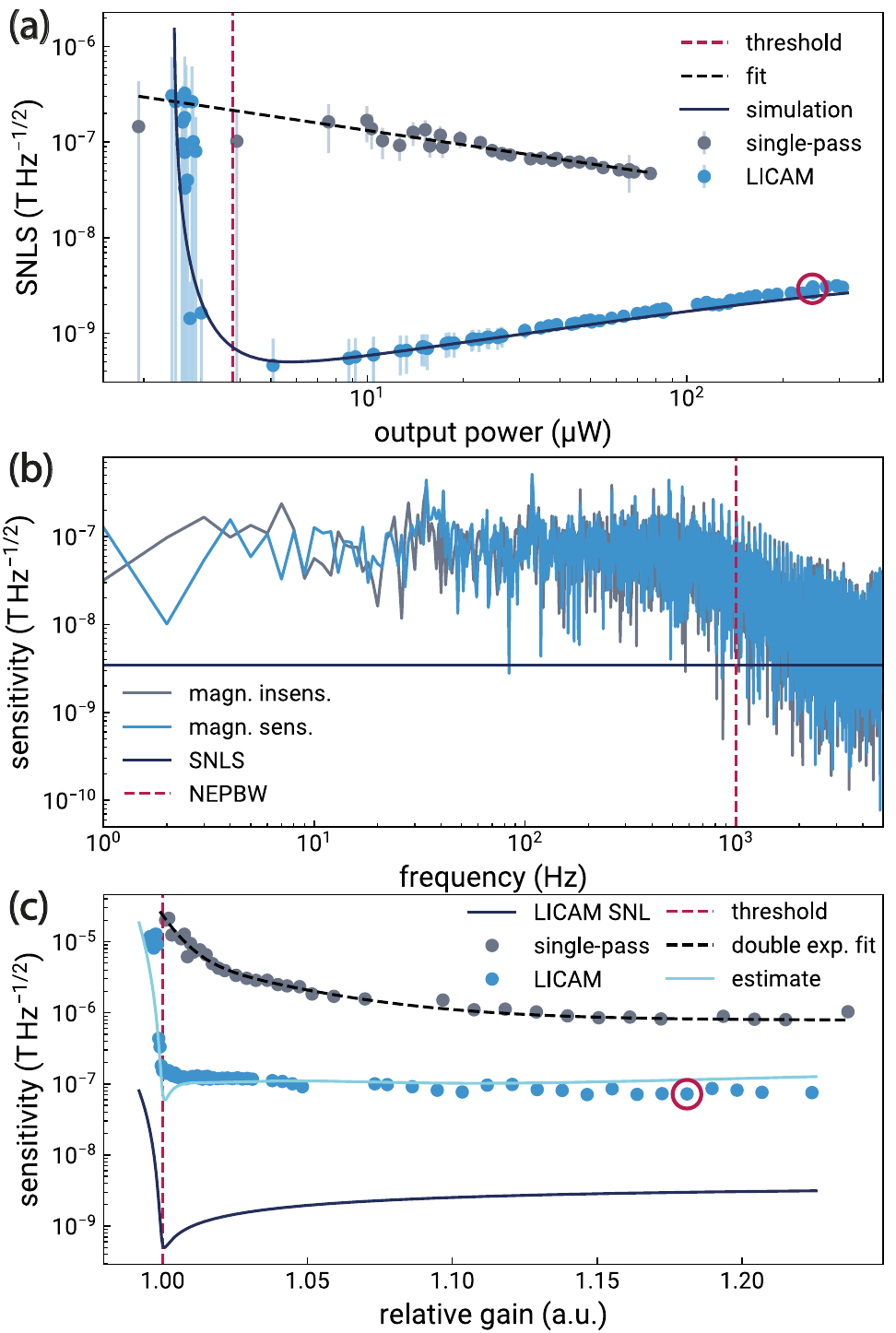}
        \caption[]{(a) Shot-noise-limited sensitivity of the single-pass (grey) and LICAM (blue) configurations. The black dashed line represents a fit according to equation (\ref{eq:snls}). The solid blue line shows the result of a simulation (see sec. \hyperref[sec:rateq]{simulation}), and the red dashed line marks the optical power at the lasing threshold. (b) Sensitivity measurement procedure: A $1\,$s long time series is recorded and Fourier transformed to obtain the noise floor for the on-resonant (blue) and off-resonant (grey) case. The noise-equivalent power bandwidth is indicated by a red dashed line, and the vertical line marks the shot-noise limit. This measurement corresponds to the circled datapoint in (a) and (c) for a relative gain of $1.18$. (c) Measured sensitivity versus gain current for the single-pass (grey) and LICAM (blue) configuration. Single-pass sensitivities were interpolated using a double-exponential decay (black dashed line) and multiplied by the simulated enhancement factor to obtain an estimate for the LICAM sensitivities (light blue line). The dark blue line marks the calculated shot-noise limit.} 
        \label{fig:3}
        \end{figure}

        From fits of the ODMR traces the shot-noise limited sensitivity (SNLS) is calculated using equation (\ref{eq:snls}). FIG. \ref{fig:3}(a) shows the result for LICAM and single-pass as a function of output power. Approaching the lasing threshold, the SNLS improves rapidly from several hundred $\text{nT}\,\text{Hz}^{-1/2}$ to $1\,\text{nT}\,\text{Hz}^{-1/2}$. The optimum value of $500(200)\,\text{pT}\,\text{Hz}^{-1/2}$ occurs slightly above the threshold at $P_\mathrm{IR}=5\,$µW. Increasing the gain further reduces the shot-noise level due to higher output power, but this improvement is offset by the simultaneous drop in spin contrast, resulting in a net increase in the SNLS. Compared with the single-pass configuration, which follows the expected $1/\sqrt{P_\mathrm{IR}}$ dependence, LICAM achieves up to a $400$-fold SNLS enhancement. We emphasize that this performance is realized at a low output power of only $5\,$µW, whereas conventional optical absorption magnetometers typically operate at power levels of several milliwatts \cite{TayefehYounesi.2025}. We measure the magnetic sensitivity of the sensor in the zero-crossing points of the ODMR traces (as shown in FIG. \ref{fig:3}(b) and (c)) by recording noise floors. FIG. \ref{fig:3}(c) shows the measured sensitivities as a function of gain current. In the single-pass configuration, the sensitivity reaches approximately $1\,\text{µT}\,\text{Hz}^{-1/2}$ above threshold --- about one order of magnitude above the shot-noise limit --- and degrades to $\sim20\,\text{µT}\,\text{Hz}^{-1/2}$ near threshold, suggesting instability caused by increased relative power fluctuations. In contrast, the LICAM sensitivity improves sharply from $\sim10\,\text{µT}\,\text{Hz}^{-1/2}$ below threshold to $<100\,\text{nT}\,\text{Hz}^{-1/2}$ above threshold, indicating that the increased power fluctuations are compensated by the contrast enhancement. The optimum sensitivity of $70(2)\,\text{nT}\,\text{Hz}^{-1/2}$ is measured at a relative threshold gain of $1.18$. We emphasize, that the measured LICAM sensitivities are about two orders of magnitude away from the projected SNLS and their discrepancy is maximized at the lasing threshold. To investigate the contribution of magnetic and non-magnetic noise we record an off-resonant noise-floor at this point. FIG. \ref{fig:3}(b) shows the on- and off-resonant noise floor. No significant difference is observed, which indicates that non-magnetic noise dominates the system. Comparing the measured sensitivities with the simulation, we find good agreement above threshold. However near threshold, the measured values exceed predictions by a factor of $2.5$ and lack the pronounced drop seen in the simulation (FIG. \ref{fig:3}(c)). This discrepancy could arise from thermal fluctuations in the diamond induced by the pump laser, or from mechanical vibrations of the diamond mount. When the diamond is inside the cavity, such fluctuations can increase the phase and amplitude noise of the cavity mode, potentially playing a more significant role near threshold. Nevertheless, an optimal sensitivity enhancement of $180$ is achieved at threshold.

        \subsection*{Towards femtotesla sensitivities with low power requirements}

\begin{figure}[ht!]	
        \centering
        \includegraphics[width=0.49\textwidth]{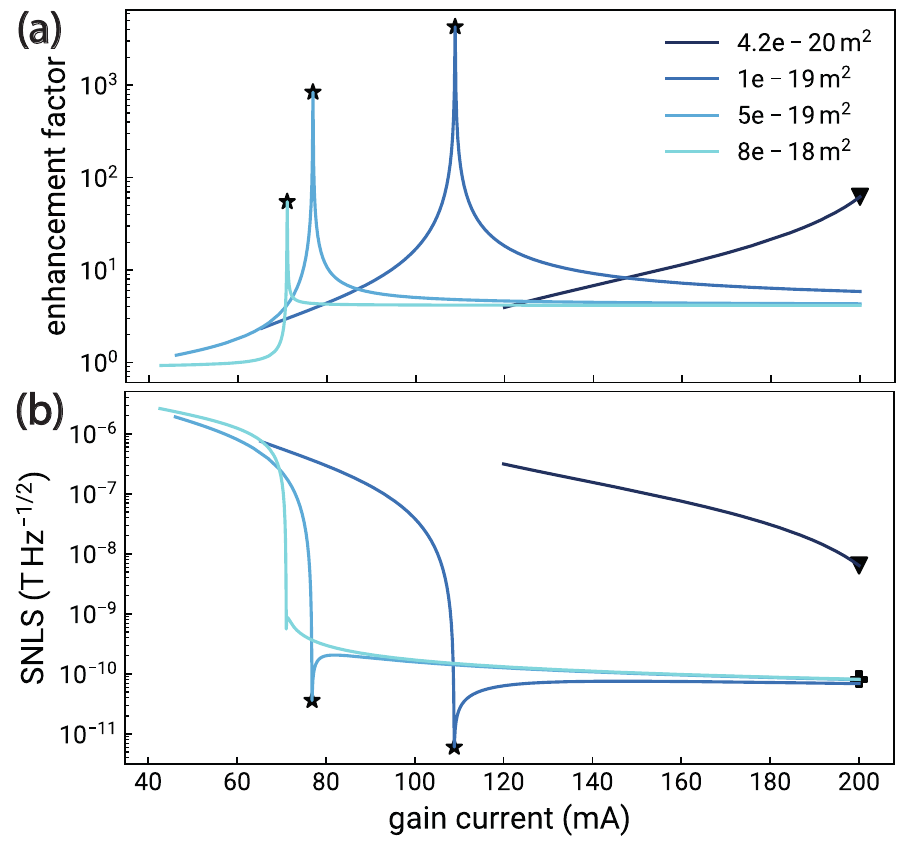}
        \caption[]{(a) Simulation of the LICAM enhancement factor $\xi(I)$ with respect to the gain current $I$ for varying differential gain $g$. (b) Simulation of the shot-noise limited sensitivity $\eta$. The black markers highlight if the optimum values were achieved at the lasing threshold (star), above the threshold at the current limit of $I_\mathrm{c}=200\,$mA (plus) or if the threshold was not reached (triangle). $R_\mathrm{f}=0.8$, $\Delta\alpha=20\,\text{m}^{-1}$. Other parameters as listed in the supplementary, TAB. I. } 
        \label{fig:4}
        \end{figure}

\begin{figure}[t!]
            \centering
            \includegraphics[width=.49\textwidth]{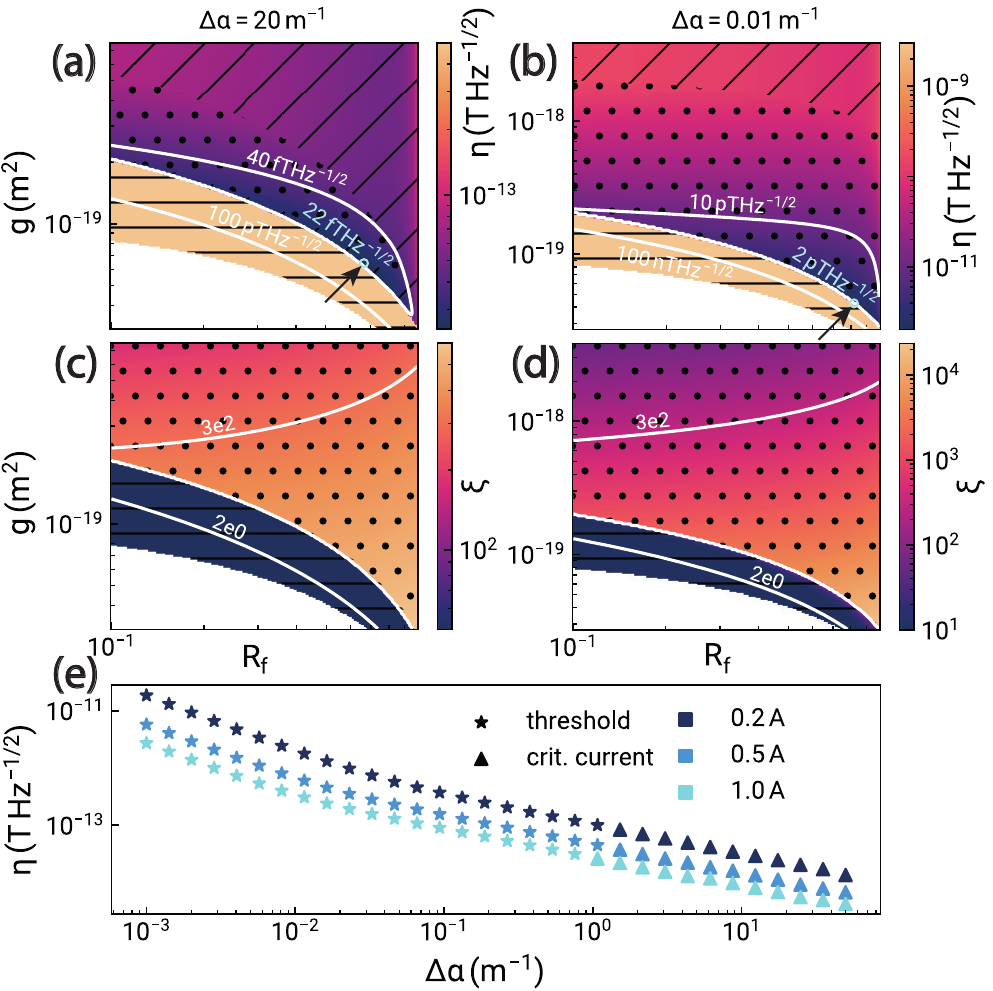}
            \caption[]{Simulated optimal values of the enhancement factor $\xi(I)$ and the SNLS $\eta(I)$ with a current limit of $I_\mathrm{c}=200\,$mA, with respect to the differential gain $g$ and the front facet reflectivity $R_\mathrm{f}$, using improved parameters (Supplementary TAB. I). (a), (b): Shot-noise limited sensitivity and (c), (d): Enhancement factor for $\Delta\alpha=20\,\text{m}^{-1}$ and $\Delta\alpha=0.01\,\text{m}^{-1}$, respectively. White contour lines indicate iso-levels, and the pale blue circles in (a) and (b) the optimum value. Hatchings indicate in which operating regime the optimum occurs: below the lasing threshold (horizontal lines), at the lasing threshold (dots), or at the current limit (diagonal lines). (e) Optimum SNLS as a function of the differential absorption constant $\Delta\alpha$ for different current limits. The marker shape indicates threshold operation (stars) or operation at the current limit (triangles).}
            \label{fig:5}
        \end{figure}

        Having established a validated model that reproduces the experimental results, we now explore the shot-noise-limited sensitivities (SNLSs) achievable assuming realistic parameters reported in the literature. Our model is based on single-mode diode-laser rate equations for coupled dynamics of the intracavity optical power $P$ and the excess carrier density $N$ \cite{Kastner.2019,Wenzel.2021}. In this model, the ECDL chip, external cavity, diamond and  lens are effectively represented as homogeneous laser cavity of length $l$. Optical gain is provided by the carrier-density-dependent gain function $G(N)$, whereas all passive optical losses are combined into one effective absorption constant $\alpha$. The ODMR signal is modeled as a MW-induced differential absorption $\Delta\alpha$, such that the off- and on-resonant steady-state cavity powers are given by $P_\alpha(I)$ and $P_{\alpha+\Delta\alpha}(I)$, respectively. The ratio of these determines an effective increase in the optical path length \cite{Scherer.2009}, from which the LICAM contrast, enhancement factor and SNLS can be calculated. The model is first constrained by fitting the measured characteristic curve (see FIG. \ref{fig:2}(a)) and is then used to extrapolate the SNLS under improved device parameters and within practical power constraints. We classify device parameters into primary parameters - those that directly improve the enhancement - and secondary parameters, which must be optimized for best performance (see supplementary for details). The most relevant parameters in the extrapolation are the differential absorption $\Delta\alpha$, the front-facet reflectivity $R_f$, and the differential gain $g$, which defines how sensitive the material gain $G(N)$ responds to changes in the carrier density. For a given set of parameters, we compute the SNLS $\eta_\mathrm{SNLS}$ and the enhancement factor $\xi$ as functions of gain current and identify their respective optima (Supplementary FIG. S5). The enhancement factor is always maximized at the lasing threshold and approaches a constant value for higher gain currents, as the gain saturates. The SNLS exhibits a local minimum at the threshold and approaches a slope proportional to $1/\sqrt{P_\mathrm{IR}}$ above it. We find that the local SNLS optimum at threshold improves with higher differential absorption constants $\Delta\alpha$ and lower differential gain $g$, albeit at the cost of higher drive currents (Supplementary FIG. S6). However, in realistic environments --- particularly for fiber-packaged, scaled-up or miniaturized devices --- power limits must be observed. This particularly hinders VECSEL-based gain chips, which require pump lasers with power levels on the order of several watts \cite{Dumeige.2019,Gottesman.2024}. In contrast, our ECDL-chip is electrically driven with a gain power consumption of $174\,$mW operated at the lasing threshold. In the following simulations, we therefore impose a gain current limit of $200\,$mA, corresponding to the maximum operating current of the ECDL chip. FIG. \ref{fig:4} presents simulations for $\eta_\mathrm{SNLS}(I)$ and $\xi(I)$ for different differential gain values with improved parameters. Both the enhancement factor and the SNLS improve as the differential gain decreases and the threshold current approaches the current limit. For larger values of $g$, the threshold is reached at much lower currents, and operation at the current limit yields a higher SNLS than operation at the threshold. We systematically investigate this behavior by calculating $\eta_\mathrm{SNLS}(I)$ and $\xi(I)$ for varying $g$ and front facet reflectivity $R_f$, identifying the optimum operating point within the current limit of $I_\mathrm{c}=200\,$mA. Each point is classified according to whether the threshold is not reached, the optimum occurs at the lasing threshold, or the optimum occurs at the current limit. The results of this simulation are shown in FIG. \ref{fig:5}. 
        
        We observe a lower limit $g(R)$, below which the system cannot reach the lasing threshold. Above this line, threshold operation yields the best performance. Even for small differential absorption constants of $\Delta\alpha=0.01\,\text{m}^{-1}$, the simulations predict sensitivities up to $2\,\text{pT}\,\text{Hz}^{-1/2}$ for realistic parameters (Supplementary TAB. I, prosp.). Increasing the differential absorption to $\Delta\alpha=20\,\text{m}^{-1}$ improves the SNLS up to $22\,\text{fT}\,\text{Hz}^{-1/2}$. For such high differential absorption constants, the spin contrast saturates at unity, resulting in moderate enhancement factors $\xi\approx600$ for $R_\mathrm{f}\rightarrow1$ compared with much larger values ($\xi\approx 2.4\times10^4$) obtained for $\Delta\alpha=0.01\,\text{m}^{-1}$. FIG. \ref{fig:5}(e) shows the optimum SNLS as a function of $\Delta\alpha$ for different current limits. In general, higher available current improves the SNLS. We find two scaling regimes: $\eta_\mathrm{SNLS}\propto \Delta\alpha^{-1}$ for small differential absorption constants $\Delta\alpha\lesssim10^{-2}\text{m}^{-1}$ and $\eta_\mathrm{SNLS}\propto \sqrt{\Delta\alpha}$ for $\Delta\alpha\gtrsim10^{-1}\text{m}^{-1}$ where the spin contrast saturates. In general, larger $\Delta\alpha$ improves the SNLS by increasing the MW-induced loss modulation, while lower $g$ enhances the threshold response to this loss. However, the minimum usable $g$ is constrained by the requirement that the cavity still reaches threshold within the available current range. Overall, our simulations indicate that femtotesla sensitivities are achievable even with the relatively small differential absorption constant present in this experiment ($\Delta\alpha\approx0.038\,\text{m}^{-1}$). The key requirements to realize such a system are sub-MHz ODMR linewidths, near-unity reflectivity of the rear facet, low passive background losses and a diode-laser design, in which the differential gain and front facet reflectivity are optimized for the given background losses and differential absorption (for more details see parameter optimization in the supplementary).

    \section*{Discussion}
    We have demonstrated laser intracavity absorption magnetometry (LICAM) for the first time using a self-sustained infrared laser diode system. By comparing the measurements with an equivalent single-pass configuration, we observe a $475$-fold enhancement in ODMR contrast and a $180$-fold improvement in magnetic sensitivity at the lasing threshold. By combining laser-threshold magnetometry with intracavity absorption spectroscopy, LICAM extends the threshold-based concept to a continuous absorption-readout scheme. In the present experiment, we do not resolve a MW-induced shift of the threshold current, instead the data is well described by a laser diode rate-equation model in which the ODMR signal arises from a MW-induced modulation of the intracavity absorption. Moreover, we show that sensitivity enhancements occur not only at the lasing threshold but also well above it. In this above-threshold regime, the contrast enhancement decreases by approximately an order of magnitude because the relative influence of an additional intracavity loss becomes weaker with increasing gain current. However, the increased laser stability above threshold improves the measured sensitivity, moving the operating point closer to the shot-noise-limited regime. Our sensor is partially integrated, current driven, and has an active region of about $(300\text{µm})^3$. With a total system footprint already below $3\times2\times2\,$cm$^3$, it can be further integrated into a compact, portable device. \\
    
    Comparing our experimental parameters with those reported in the literature reveals substantial opportunities for optimization. Replacing the current cross-configuration of the infrared and green pump beams with a collinear geometry \cite{TayefehYounesi.2025} would significantly increase the differential absorption constant. Driving the hyperfine transitions could further steepen the ODMR slope. Using an impedance-matched antenna \cite{Opaluch.2021} and a doubly resonant cavity design \cite{Dumeige.2013} for the green pump light could significantly reduce the power requirements, which are much larger than the electrical gain power of $174\,$mW in the current configuration ($P_\mathrm{MW}=1.6\,\text{W},P_\mathrm{pump}=1.4\,\text{W}$).  The DBR grating can be designed to match the NV$^-$ singlet resonance, eliminating the need for temperature control. In a next step, the diamond can be integrated directly on-chip together with the output coupler, removing the lens and thereby reducing additional scattering losses in the cavity. This integration would enable a compact, millimeter-scale sensor architecture. Our simulations indicate that, with optimized yet realistic control and chip parameters, femtotesla shot-noise-limited sensitivities are achievable for moderate differential absorption constants above $10^{-1}\,\text{m}^{-1}$. Specifically, for $\Delta\alpha=20\,\text{m}^{-1}$ we predict a limit of $22\,\mathrm{fT}\,\mathrm{Hz}^{-1/2}$.  For comparison, recent infrared-absorption-based NV magnetometers report projected shot-noise-limited sensitivities of \(26.6\,\mathrm{pT}\,\mathrm{Hz}^{-1/2}\) and \(<400\,\mathrm{fT}\,\mathrm{Hz}^{-1/2}\) for VECSEL- \cite{Gottesman.2024} and MECSEL-based \cite{schall.2026} laser threshold magnetometry, and \(5\,\mathrm{pT}\,\mathrm{Hz}^{-1/2}\) for optimized single-pass infrared absorption \cite{TayefehYounesi.2025}. The best reported photoluminescence-collection-based NV ensemble magnetometer reaches calculated photon-shot-noise limits of \(260\,\mathrm{fT}\,\mathrm{Hz}^{-1/2}\) for broadband near-DC Ramsey magnetometry and \(90\,\mathrm{fT}\,\mathrm{Hz}^{-1/2}\) for narrowband AC Hahn-echo magnetometry \cite{Barry.2024}. Thus, the predicted LICAM shot-noise limit compares favorably with these reported shot-noise limits, while retaining an electrically driven and partially integrated infrared probe architecture. \\
    Experimentally, our measured sensitivity is not limited by magnetic noise but rather by system noise, most likely dominated by power fluctuations and mechanical vibrations of the external cavity. Additional stabilization schemes or the above discussed chip-integration could mitigate these effects. We also emphasize that LICAM can be operated with pulse sequences, enabling quantum sensing protocols that can further boost sensitivity. Finally, LICAM is not restricted to NV$^-$-based sensors but can be extended to a broad class of optically interactive quantum sensors, including OPMs that rely on absorption or polarization rotation.  \\
    A key challenge that remains is increased relative power fluctuations near the lasing threshold. To address this limitation, we propose multimode LICAM, which --- analogous to multimode intracavity absorption spectroscopy \cite{Baev.1999,Fjodorow.2024}  --- can operate stably far above the lasing threshold. In contrast to single-mode operation, multimode LICAM would be inherently robust against broadband cavity losses and fluctuations and could enable balanced photodetection schemes \cite{Hobbs.1991, Acosta.2010_broadband, TayefehYounesi.2025}. Noise mitigation strategies and multimode LICAM are briefly discussed in the supplementary.\\
    Realizing these improvements would enable magnetometry performance approaching the shot-noise limit, opening novel opportunities for applications in biomedical sensing, material sciences and beyond. 

    \section*{Methods}
    \label{sec:mm}

    \subsection*{Experiment}

    We use a high pressure high temperature (HPHT) diamond with an NV$^-$ density of approximately $7\,$ppm (more details on sample fabrication are stated in the supplementary). Both ECDL edge facet and diamond facets are antireflection-coated for the target wavelength of $1042\,$nm. A tunable distributed Bragg reflector (DBR) selects the  $1042\,$nm infrared resonance from the broad emission spectrum of the current-driven InGaAs gain medium. A lens collimates the edge emission from the gain chip, which then passes through the diamond, behind which a coupling mirror ($R_f=0.9$) closes the external cavity. By tuning the gain current and measuring the optical output power of the cavity we obtain the characteristic power-current curve of a single-mode diode laser (see FIG. \ref{fig:2} (a)), with a lasing threshold of $I_\mathrm{th}=116\,$mA. The diamond is pumped perpendicularly by a $532\,$nm green laser, defining an overlap region with the infrared beam that constitutes the active sensing volume. The diamond is cut along the $[111]$-direction and oriented such that both the infrared and the green laser are linearly polarized in the $[111]$-plane, maximizing the optical absorption cross-sections. A copper antenna delivers a microwave (MW) signal driving the NV$^-$ spin transitions. We apply sinusoidal frequency modulation at $8\,$kHz with a modulation depth of $4.5\,$MHz to enable lock-in detection and optimize for the pump and MW powers. Sweeping the carrier frequency at a fixed gain current yields ODMR spectra (Supplementary FIG. S2(a)). A ring magnet is used to split the spin resonances and isolate the $([111],-1_{\mathrm{m}_\mathrm{S}})$-resonance which is used for all measurements. We vary the gain current along the characteristic laser curve and monitor the optical output power while simultaneously recording ODMR spectra, from which the spin contrast is extracted by fitting the derivative of a gaussian resonance profile. Further details can be found in the supplementary. From the fits, we extract the ODMR linewidth which, together with the spin contrast and the optical output power $P_\mathrm{IR}$, allows to calculate the SNLS of the LICAM-sensor using the standard expression \cite{Dreau.2011, Acosta.2010_broadband}:

        \begin{equation}
            \label{eq:snls}
            \eta_{\mathrm{SNLS}} = \sqrt{\frac{e}{8\ln2}}\frac{1}{\gamma}\frac{\Delta\nu}{C}\sqrt{\frac{E_\mathrm{IR}}{P_\mathrm{IR}}},
        \end{equation}
    where $\gamma$ is the electron gyromagnetic ratio, $C$ the spin contrast, $\Delta\nu$ the ODMR linewidth, and $E_\mathrm{IR}$ the infrared photon energy. To determine experimental enhancement factors we repeat the measurements in a single-pass configuration, where the diamond is placed outside the cavity and thus does not contribute optical feedback. To measure the magnetic sensitivity, the carrier frequency is tuned to the ODMR zero-crossing point $\nu_0=2.7435\,$GHz for each gain current (see FIG. \ref{fig:2}(b) and (c)). A $1\,$s time trace is recorded at a $10\,$kHz sampling rate and Fourier transformed to obtain the amplitude spectral density (ASD), as shown in FIG. \ref{fig:3}.(b). The AC magnetic sensitivity is calculated as the average noise floor in the $1\,$Hz to $1\,$kHz bandwidth. To distinguish magnetic from non-magnetic noise sources, the measurement is repeated with an off-resonant carrier frequency $\nu_\mathrm{off}=2.65\,$GHz. We interpolate the single-pass sensitivity using a double-exponential decay and scale it by the simulated LICAM enhancement factor to estimate the LICAM sensitivity. The LICAM SNLS in FIG. \ref{fig:3}(c) is extracted from the simulation $\eta_{\mathrm{SNLS}}(P_\mathrm{IR})$ shown in (a) together with the fit of the characteristic curve $P_\alpha(I)$.
    
    \subsection*{Simulation}
    \label{sec:rateq}
    
    To validate our experimental observations of intracavity enhancement of contrast and sensitivity near the lasing threshold, we employ the following rate-equation model for a single-mode diode laser including spontaneous emission for the average optical power in the cavity $P$ and the excess carrier density $N$ \cite{Kastner.2019,Wenzel.2021}:

    \begin{equation}
    \label{eq:sde}
    \begin{cases}
    \begin{aligned}
    &\frac{1}{v_\mathrm{g}}\frac{dP}{dt} = \left[\Gamma G(N) - \alpha - \Gamma\sigma N \right]P + \frac{Kv_{\mathrm{g}}\hbar\omega\Gamma r_{\mathrm{sp}}(N)}{l} \\
    \\
    &\frac{dN}{dt} = \frac{I}{edWl} - R(N) - \frac{\Gamma G(N)P}{\eta_\mathrm{i} dW\hbar\omega}
    \end{aligned}
    \end{cases}
    \end{equation}
    Here, $v_\mathrm{g}=c/n_\mathrm{g}$ is the group velocity, $\alpha$ the total cavity loss absorption constant, $\sigma$ the cross-section for free carrier absorption, $I$ the gain current, $\omega=2\pi c/\lambda$, $\omega$ and $\lambda$ the angular frequency and wavelength, $e$ the elementary charge, $\hbar\omega$ the photon energy, $d$ the thickness of the active region, $W$ the width of the active region, $l$ the length of the active region, and $\eta_\mathrm{i}$ the internal efficiency. The optical confinement factor $\Gamma$ describes the spatial overlap of the optical mode with the electrically pumped gain region and scales both stimulated emission and free-carrier absorption. The Petermann K-factor $K$ accounts for excess spontaneous emission noise into the lasing mode caused by non-ideal mode properties \cite{Petermann.1979, Siegman.1989}. We use $K=1$, assuming an ideal mode. By linearizing the gain $G(N)$, the spontaneous emission $r_{\mathrm{sp}}(N)$, and the recombination rate $R(N)$ around the threshold carrier density $N_\mathrm{th}$, equation (\ref{eq:sde}) can be solved analytically for $P$ in the steady state (Supplementary equation (8)). This yields the average cavity power $P_\alpha(I)$ at fixed absorption $\alpha$ as a function of gain current. This linearization substantially reduces computational cost, allowing efficient fitting routines and large-scale parameter sweeps that would otherwise require numerical integration of equation (\ref{eq:sde}). The optical power exiting the cavity through the front facet and detected by the photodiode is given by \cite{Coldren.2012, Kastner.2019}:

    \begin{equation}
        P^{(\mathrm{out})}_\alpha(I) = P_\mathrm{sp}-\frac{P_\alpha(I)}{2}\ln\left(R_\mathrm{r}R_\mathrm{f}\right) \left[ 1+\frac{1-R_\mathrm{r}}{1-R_\mathrm{f}}\sqrt{\frac{R_\mathrm{f}}{R_\mathrm{r}}}\right]^{-1} ,
        \label{eq:Pout}
    \end{equation}
    with $R_\mathrm{r}$ and $R_\mathrm{f}$ denoting the rear and front reflectivities of the cavity, respectively.
    We account for background spontaneous emission $P_\mathrm{sp}$ into non-lasing modes that are still detected by the photodiode and contribute to the single-pass optical depth $\tau_\mathrm{r}$. By combining equations (\ref{eq:sde}) and (\ref{eq:Pout}), we extract $\eta_\mathrm{i}$, $P_\mathrm{sp}$ and $N_\mathrm{th}$ from a fit to the characteristic curve shown in FIG. \ref{fig:2}(a). We define $\Delta\alpha$ as the MW-induced differential absorption constant. On MW resonance, the total absorption constant is therefore $\alpha+\Delta\alpha$. With this definition, $\Delta\alpha$  can be derived from the measured single-pass spin contrast $C_\mathrm{sp}$, assuming that the optical depth of the overlap region on resonance is equivalent in the LICAM configuration. This enables calculation of the on-resonant LICAM output power $P_{\alpha+\Delta\alpha}(I)$ (see supplementary). Following \cite{Scherer.2009}, the intracavity enhancement can be described in terms of an effective optical path length that increases the optical depth:
    
    \begin{equation}
        \label{eq:depth}
        \tau_\mathrm{eff}(I) = \tau_\mathrm{r} + \ln{\left(\frac{P_{\alpha}(I)}{P_{\alpha+\Delta\alpha}(I)}\right)}.
    \end{equation}
    Using equation (\ref{eq:depth}), we calculate the LICAM contrast $C(I)=1-\exp\lbrace-\tau_\mathrm{eff}(I)\rbrace$, the enhancement factor $\xi(I)=\tau_\mathrm{eff}(I)/\tau_\mathrm{r}$, and the LICAM shot-noise-limited sensitivity using $P_{\alpha+\Delta\alpha}(I)$ and $C(I)$ in equation (\ref{eq:snls}). The results of these simulations are shown in FIGs. \ref{fig:2}.(d) and \ref{fig:3}.(a). We find excellent agreement with the experimental data, confirming that the rate-equation model (equation (\ref{eq:sde})), together with the applied linearization, adequately describes the LICAM-enhancement of both ODMR contrast and magnetic sensitivity. All simulation parameters are listed in the supplementary, TAB. I.

    \section*{Acknowledgement}
    The authors acknowledge Viviana Villafane for annealing the samples, Peter Ressel and Uwe Sprengler for AR coating the side facets of the sample, Christof Zink for measuring the emission profile of the ECDL chip, and Tommaso Pregnolato for the tri-acid boiling. We thank Sascha Neinert and Kirti Vardhan for a fruitful discussion related to OPMs.

    This project is funded by the German Federal Ministry of Research, Technology and Space (BMFTR) within the “DiNOQuant” project (No. 13N14921), the European Research Council within the ERC Starting Grant “QUREP” (No. 851810), and the Einstein Research Unit “Perspectives of a quantum digital transformation: Near-term quantum computational devices and quantum processors”.

    \vspace{1em}

    J.W., F.P. and T.S. developed the concept of the LICAM- sensor. J.W. and F.P. designed the experiment. J.W. and A.P. performed the measurements, J.W. analyzed the data. A.P. performed and analyzed the measurements for the sensitivity optimization and the sensor demonstration. H.W. provided the rate equation model, simulation parameters for the ECDL-chip and supported the simulations done by J.W. H.C. und A.K. provided the ECDL-chip. W.K. carried out the electron irradiation of the diamond sample. J.M.B. developed the software platform for the experiments and provided helpful discussions of the experimental data. J.W., A.P. and T.S. wrote the manuscript. All authors contributed to the writing.

    \bibliography{refz}



\end{document}


\makeatletter
    \def\@caption#1[#2]#3{%
      \par\addcontentsline{\csname ext@#1\endcsname}{#1}{%
      \protect\numberline{\csname the#1\endcsname}{#2}}%
      \begingroup
        \@parboxrestore
        \leftskip=0pt
        \rightskip=\@flushglue
        \parindent=0pt
        \parfillskip=0pt
        \noindent
        {\bfseries \csname fnum@#1\endcsname:} #3\par
      \endgroup}
    \makeatother

    \input{supplementary/title_sup}
    \maketitle

    \input{supplementary/supplementary}

    \bibliography{supplementary/refz}

%% file: preambel.tex
\usepackage{graphicx}
\usepackage[usenames,dvipsnames]{xcolor}
\usepackage{amssymb,amsfonts,amsmath}
\usepackage[colorlinks=true,citecolor=Cerulean,linkcolor=RubineRed,urlcolor=Cerulean]{hyperref}
\usepackage{epstopdf}
\usepackage{bm}
\usepackage{bbold}
\usepackage{upgreek}
\usepackage[inline]{enumitem}
\usepackage{booktabs}
\usepackage{multirow}
\usepackage[normalem]{ulem}
\usepackage{IEEEtrantools}
\usepackage{subfigure}
\usepackage{subcaption}

\usepackage{float}

\usepackage{afterpage}

\usepackage{verbatim}

\usepackage{lipsum, babel}
\usepackage{sidecap}
\usepackage{hyperref}
\hypersetup{
    colorlinks=true,
    linkcolor=magenta,
    filecolor=magenta,      
    urlcolor=cyan,
    pdftitle={Overleaf Example},
    pdfpagemode=FullScreen,
    }
\usepackage{xr}
\makeatletter
\newcommand*{\addFileDependency}[1]{
  \typeout{(#1)}
  \@addtofilelist{#1}
  \IfFileExists{#1}{}{\typeout{No file #1.}}
}
\makeatother

\newcommand{\ket}[1]{\left|#1\right\rangle}

\usepackage{txfonts}
\usepackage{latexsym}

\graphicspath{ {./figz/} }
\newlist{custominlist}{enumerate*}{1}
\setlist[custominlist]{label=(\roman*)}

%% file: title.tex
\title{Laser intracavity absorption magnetometry for optical quantum sensing}

\author{J. M. Wollenberg}
\affiliation{Department of Physics, Humboldt-Universit\"{a}t zu Berlin, Newtonstr. 15, 12489 Berlin, Germany}

\author{F. Perona}
\affiliation{Ferdinand-Braun-Institut (FBH), Gustav-Kirchhoff-Str. 4, 12489 Berlin, Germany}

\author{A. Palaci}
\affiliation{Department of Physics, Humboldt-Universit\"{a}t zu Berlin, Newtonstr. 15, 12489 Berlin, Germany}

\author{H. Wenzel}
\affiliation{Ferdinand-Braun-Institut (FBH), Gustav-Kirchhoff-Str. 4, 12489 Berlin, Germany}

\author{H. Christopher}
\affiliation{Ferdinand-Braun-Institut (FBH), Gustav-Kirchhoff-Str. 4, 12489 Berlin, Germany}

\author{A. Knigge}
\affiliation{Ferdinand-Braun-Institut (FBH), Gustav-Kirchhoff-Str. 4, 12489 Berlin, Germany}

\author{W. Knolle}
\affiliation{Leibniz-Institut für Oberflächenmodifizierung e.V., 04318 Leipzig, Germany}

\author{J. M. Bopp}
\affiliation{Department of Physics, Humboldt-Universit\"{a}t zu Berlin, Newtonstr. 15, 12489 Berlin, Germany}
\affiliation{Ferdinand-Braun-Institut (FBH), Gustav-Kirchhoff-Str. 4, 12489 Berlin, Germany}

\author{T.~Schr\"{o}der}
\email[Corresponding author: ]{tim.schroeder@physik.hu-berlin.de}
\affiliation{Department of Physics, Humboldt-Universit\"{a}t zu Berlin, Newtonstr. 15, 12489 Berlin, Germany}
\affiliation{Ferdinand-Braun-Institut (FBH), Gustav-Kirchhoff-Str. 4, 12489 Berlin, Germany}

\date{\today}

%% file: abstract_new2.tex
\begin{abstract}
    Intracavity absorption spectroscopy (ICAS) is a well-established technique for detecting weak absorption signals with ultrahigh sensitivity. Here, we extend this concept to magnetometry using nitrogen-vacancy (NV) centers in diamond. We introduce laser intracavity absorption magnetometry (LICAM), a concept that is in principle applicable to a broader class of optical quantum sensors, including optically pumped magnetometers. Using an electrically driven, edge-emitting diode laser that operates self-sustainably, we show that LICAM enables highly sensitive magnetometers operating under ambient conditions. Near the lasing threshold, we achieve a 475-fold enhancement in optical contrast and a 180-fold improvement in magnetic sensitivity compared with a conventional single-pass geometry. The experimental results are accurately described by a rate-equation model for single-mode diode lasers. From our measurements, we determine a projected shot-noise-limited sensitivity in the $\mathrm{pT}\,\mathrm{Hz}^{-1/2}$ range and simulate that, with realistic device improvements, shot-noise limited sensitivities down to the $\mathrm{fT}\,\mathrm{Hz}^{-1/2}$ scale are attainable.
\end{abstract}

%% file: supplementary/title_sup.tex
\title{Supplementary Material to Laser Intracavity Absorption Magnetometry for Optical Quantum Sensing}

\author{J. M. Wollenberg}
\affiliation{Department of Physics, Humboldt-Universit\"{a}t zu Berlin, Newtonstr. 15, 12489 Berlin, Germany}

\author{F. Perona}
\affiliation{Ferdinand-Braun-Institut (FBH), Gustav-Kirchhoff-Str. 4, 12489 Berlin, Germany}

\author{A. Palaci}
\affiliation{Department of Physics, Humboldt-Universit\"{a}t zu Berlin, Newtonstr. 15, 12489 Berlin, Germany}

\author{H. Wenzel}
\affiliation{Ferdinand-Braun-Institut (FBH), Gustav-Kirchhoff-Str. 4, 12489 Berlin, Germany}

\author{H. Christopher}
\affiliation{Ferdinand-Braun-Institut (FBH), Gustav-Kirchhoff-Str. 4, 12489 Berlin, Germany}

\author{A. Knigge}
\affiliation{Ferdinand-Braun-Institut (FBH), Gustav-Kirchhoff-Str. 4, 12489 Berlin, Germany}

\author{W. Knolle}
\affiliation{Leibniz-Institut für Oberflächenmodifizierung e.V., 04318 Leipzig, Germany}

\author{J. M. Bopp}
\affiliation{Department of Physics, Humboldt-Universit\"{a}t zu Berlin, Newtonstr. 15, 12489 Berlin, Germany}
\affiliation{Ferdinand-Braun-Institut (FBH), Gustav-Kirchhoff-Str. 4, 12489 Berlin, Germany}

\author{T.~Schr\"{o}der}
\email[Corresponding author: ]{tim.schroeder@physik.hu-berlin.de}
\affiliation{Department of Physics, Humboldt-Universit\"{a}t zu Berlin, Newtonstr. 15, 12489 Berlin, Germany}
\affiliation{Ferdinand-Braun-Institut (FBH), Gustav-Kirchhoff-Str. 4, 12489 Berlin, Germany}

\date{\today}

%% file: supplementary/supplementary.tex
\setcounter{figure}{0}
\renewcommand{\thefigure}{S\arabic{figure}}

\label{sup}
    \section*{experimental Setup}

        \begin{figure*}
            \centering
            \includegraphics[width=.99\textwidth]{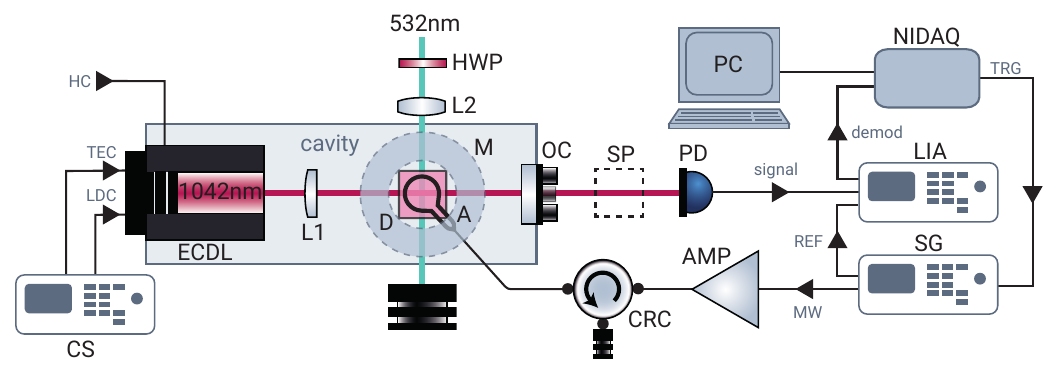}
            \caption[LTM setup]{Schematic of the LICAM setup. An ECDL chip (ECDL) emits light at $1042\,$nm into an external cavity formed by the diamond (D) and a coupling mirror (OC). The cavity output is detected by a photodetector (PD), while the diamond is side-pumped by a $532\,$nm green laser. A MW antenna (A) delivers a MW signal generated by a signal generator (SG) to drive the spin resonances. The MW signal is frequency-modulated, and the photodetector output is demodulated using a lock-in amplifier (LIA). A ring-shaped magnet (M) positioned above the diamond provides a magnetic bias field that splits the spin resonances.}
            \label{fig:sup_1}
        \end{figure*}

         The LICAM setup is illustrated schematically in FIG. \ref{fig:sup_1}. An InGaAs-based ECDL chip (ECDL) provides broadband optical gain with an emission spectrum covering $900-1100\,$nm. A DBR grating selects a narrow spectral component from this emission that can be temperature-tuned to $1042\,$nm by supplying a heating current (HC), as shown in FIG. \ref{fig:sup_2}(b). The ECDL chip is driven by an \textit{Arroyo Instruments 6310 ComboSource} (CS) which supplies the laser diode current (LDC) and stabilizes the case temperature at $25^\circ$C (TEC) using a \textit{CUI Devices CP303385H} Peltier element. The edge emission is collimated with a $f=1.45\,$mm lens (L1) and guided through the diamond sample (D). To dissipate heat from the pump laser, the sample mount is attached to a heat sink that is temperature-stabilized at $22^\circ$C using another Peltier element controlled by a \textit{Thorlabs TED200C} thermoelectric temperature controller.
        A coupling mirror (OC) with $R=0.9$ closes the external cavity. The mirror is mounted on a \textit{POLARIS-K1S2P} piezoelectric mirror mount, which is positioned on a \textit{Thorlabs NF15AP25} single-axis piezo stage for cavity alignment. The diamond is pumped perpendicularly to the IR path using a \textit{Cobolt Samba} laser operating at $532\,$nm with a power of $1.4\,$W. A $f=150\,$mm lens (L2) focuses the beam to a diameter of approximately $300\,$µm, matching the IR beam size. A half-waveplate (HWP) adjusts the pump polarization relative to the laboratory's $xy$-frame. A ring-shaped neodymium magnet (M) is placed above the diamond to lift the degeneracy of the ODMR resonances. The optical output of the cavity is detected with a \textit{LaserComponents IG17} photodetector (PD). A copper antenna (A) positioned on top of the diamond delivers the MW signal that drives the spin resonances. The MW signal is generated by a \textit{ Rohde $\&$ Schwarz SMB100B} signal generator (SG) and amplified with a \textit{Mini-Circuits ZHL-16W-43-S+} amplifier (AMP). A \textit{JQL Electronic JCC2300T3600S2R1} circulator is used to terminate reflections from the antenna. The photodetector output is demodulated with a \textit{Zurich Instruments MFLI} lock-in amplifier (LIA), which uses a reference signal (REF) from the signal generator. The demodulated signal is sampled by a \textit{NationalInstruments} data acquisition unit (NIDAQ), synchronized via a trigger signal (TRG) from the signal generator. All measurement routines are controlled by a laboratory computer (PC) running the \textit{DynExp} software \cite{Bopp.2024}. For measurements in the single-pass (SP) configuration, the diamond is placed outside of the cavity.

         \subsection*{Sample fabrication}\label{sec:fabrication}

        The sample is a \textit{CD1411} HPHT diamond from \textit{Sumitomo} with an initial nitrogen concentration of approximately $200\,$ppm. It has a cubic shape with dimensions of $1.4\,$mm$\, \times \,1.4\,$mm$\,\times\,1\,$mm and is cut along the $[111]$-direction. Electron irradiation was performed with $7\,$MeV and a fluence of $2.2\times10^{18}\text{cm}^{-2}$. Subsequent vacuum annealing at temperatures up to $1100\,^{\circ}C$ was carried out over a period of $60\,$ hours. Reference measurements indicate a NV$^-$ concentration of approximately $7\,$ppm. The inhomogeneous spin-dephasing time, measured in a Ramsey-type experiment, was found to be $96(11)\,$ns. To minimize optical scattering losses, the crystal facets were polished to a surface roughness below $3\,$nm RMS. The sample was then cleaned in a tri-acid boiling process, which led to oxygen-terminated surfaces. Two opposing facets were subsequently antireflection-coated with a $\text{Ta}_2\text{O}_5/\text{SiO}_2$ layer, yielding a reflectivity of $2.84(2)\times 10^{-3}\,\%$ at $1042\,$nm.
        
        \subsection*{Contrast calculation}
        \label{sup:contrast_calc}

        For the contrast calculations discussed in the Materials and Methods
section of the main text, the demodulated signal $S_\mathrm{dem}$ of the $(111,m_S=-1)$-peak is fitted assuming the derivative of a Gaussian resonance profile:

        \begin{equation}
            S_\mathrm{dem}(\nu) = A \exp\left\lbrace-\frac{{(\nu-\nu_0)^2}}{{2 \sigma^2}}\right\rbrace\frac{(\nu-\nu_0)}{\sigma^3\sqrt{2\pi}},
        \end{equation}
        where $A$ is the signal amplitude, $\nu$ the MW frequency, $\nu_0$ the zero-crossing frequency and $\sigma$ the linewidth parameter. Following \cite{Sturner.2021},\cite{Bopp.2025}, the ODMR contrast $C$ is obtained from the integrated signal $S_\mathrm{int}=\left\vert \int_{-\infty}^{\nu_0}\mathrm{d}\nu\,S_\mathrm{dem}(\nu)
        \right\vert$ as: 

        \begin{equation}
        \label{eq:contrast}
            C = \frac{S_\mathrm{int}}{2 I_\mathrm{S} \Delta f Z_\mathrm{LIA}},
        \end{equation}
        where $\Delta f$ is the modulation depth, $I_\mathrm{S}$ the average signal current and $Z_\mathrm{LIA}=50\,\Omega$ the input impedance of the LIA. During the measurement, $10\%$ of the cavity output power was directed to a secondary photodiode via a beamsplitter to record a reference signal on an oscilloscope. Because the ODMR resonance is symmetric around its center frequency, it does not affect the average signal level. Using a calibration of this reference signal, the average signal current on the main photodiode (connected to the LIA) can be reconstructed. The integrated signal $S_\mathrm{int}$ is then evaluated from fit parameters according to equation (\ref{eq:contrast}).

        \begin{figure}
            \centering
            \includegraphics[width=.49\textwidth]{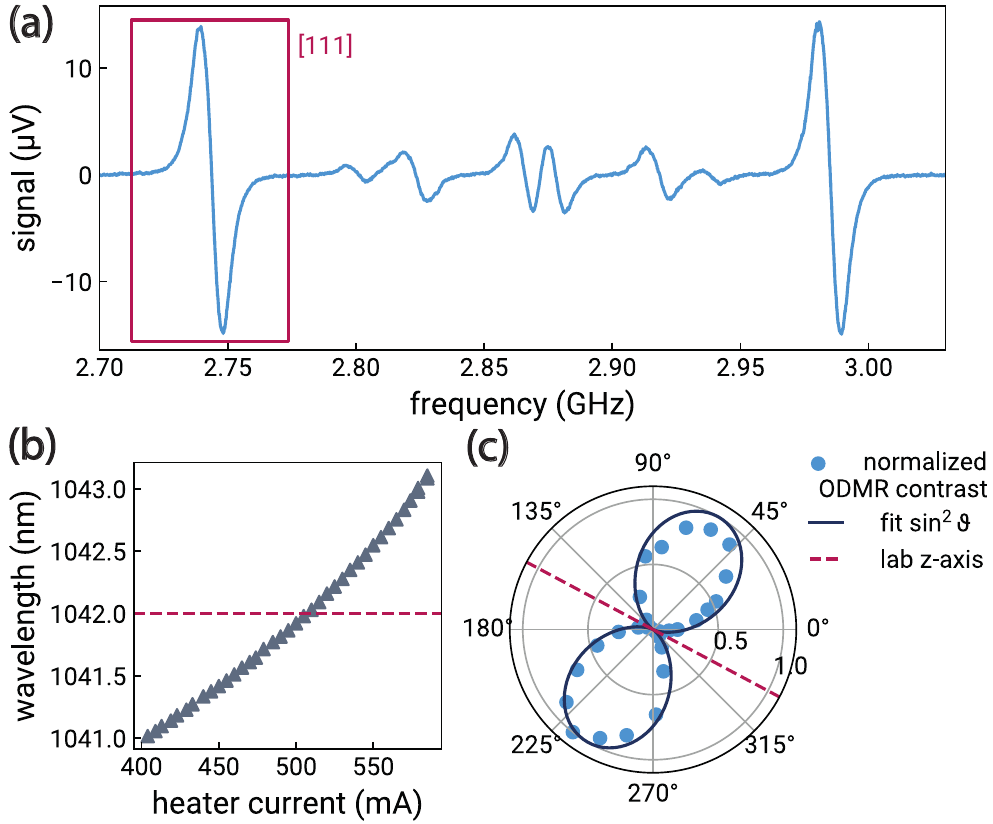}
            \caption[]{(a) ODMR spectrum of the LICAM sensor. The IR and pump polarizations are optimized to align with the dipole plane of the $[111]$ resonance (red rectangle) maximizing the spin contrast. (b) The ECDL emission wavelength can be tuned to $1042\,$nm by adjusting the heating current. (c) ODMR contrast of the $([111],-1_{\mathrm{m}_\mathrm{S}})$ resonance as a function of the pump-laser polarization angle.}
            \label{fig:sup_2}
        \end{figure}

        \section*{Sensitivity optimization}
        \ref{sup:opt}

We perform a first-order optimization of several experimental parameters to improve the magnetic sensitivity of the sensor. This includes the polarization of the green pump laser, the pump and MW powers and the LIA settings.

\subsection*{Pump and probe polarization}
The optical transitions between the NV$^-$ triplet and singlet state exhibit a dual-dipole polarization dependence. The optical absorption cross-section is maximized when the polarization vector lies in the plane orthogonal to the NV axis \cite{Acosta.2010c, peng2024}. A larger optical cross-section leads to more efficient pumping and thus a higher spin contrast. The diamond sample is cut along the $[111]$ axis, which coincides with one of the four possible NV orientations. The sample is positioned such that $[111]$ aligns with the laboratory $z$-axis. This configuration favors the edge emission of the ECDL chip, which is linearly polarized in the laboratory $xy$-plane. The magnetic bias field is also predominantly aligned along the $[111]$ direction, allowing its resonance to be isolated in the ODMR spectrum. A half-waveplate is used to rotate the polarization of the pump laser within the $yz$-plane. FIG. \ref{fig:sup_2}(c) shows the dipole absorption characteristic of the $([111],-1_{\mathrm{m}_\mathrm{S}})$ resonance as a function of the pump polarization angle. In this configuration, the ODMR contrast is maximized when the pump polarization lies in the dipole plane ($xy$).

\subsection*{Pump and microwave power}

\begin{figure*}
            \centering
            \includegraphics[width=.99\textwidth]{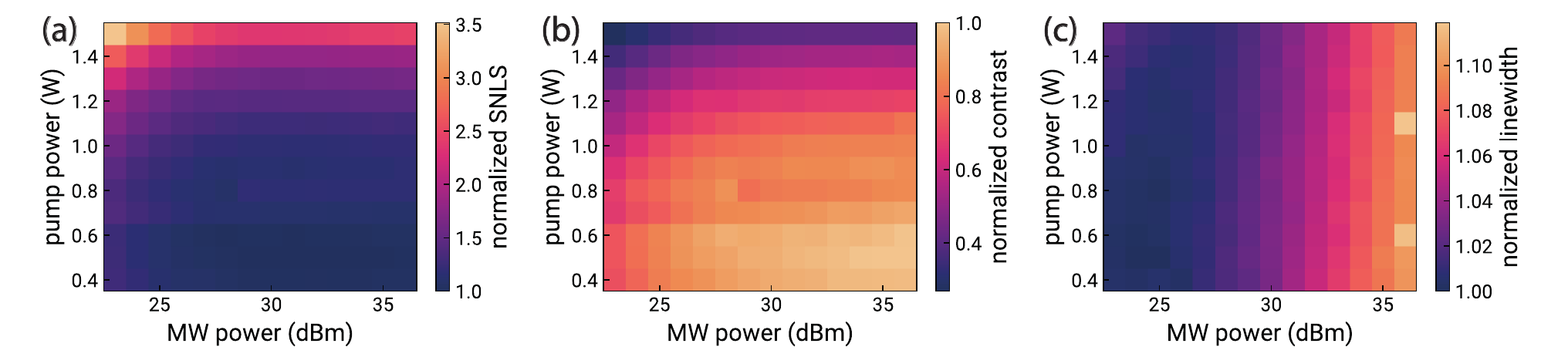}
            \caption[opt]{SNLS (a), ODMR contrast (b) and ODMR linewidth (c) versus optical pump power and MW input power. The SNLS exhibits a global optimum for $P_\mathrm{pump}=1.4\,$W and $P_\mathrm{MW}=32\,$dBm.}
            \label{fig:sup:opt}
\end{figure*}

To determine the optimal power levels of the pump laser and microwave field, we record ODMR spectra of the $([111],-1_{\mathrm{m}_\mathrm{S}})$ resonance for varying combinations of pump and MW powers. The ODMR contrast and linewidth are extracted from fits to the resonance profiles, which allows to calculate the SNLS from the contrast-to-linewidth ratio. The results are shown in FIG. \ref{fig:sup:opt}, which indicates that the SNLS reaches a global optimum for a pump power of 1.4 W and a MW input power of 32 dBm. As illustrated in FIG. \ref{fig:sup:opt}(b), the contrast increases with both pump and MW power, while the linewidth (see FIG. \ref{fig:sup:opt}(c)) broadens predominantly due to power broadening at higher MW powers. During these measurements, the LIA modulation depth was set to $f_{\text{dev}} = 6$ MHz, corresponding to roughly half the average linewidth.

\subsection*{Lock-in amplifier settings}

\begin{figure*}
            \centering
            \includegraphics[width=.99\textwidth]{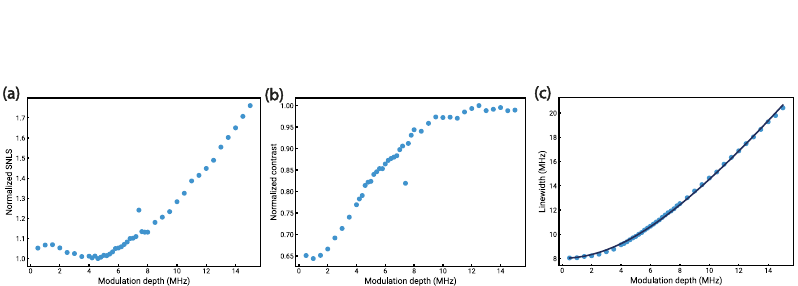}
            \caption[mod depth]{SNLS (a), ODMR contrast (b) and ODMR linewidth (c) versus modulation depth. The SNLS exhibits a global optimum at $f_\mathrm{dev}=4.5\,$MHz. A fit to the linewidth (solid line) using equation (\ref{eq:sup:lw_intr}) yields an intrinsic linewidth of $\Delta\nu_{\text{intr}}=8.02(3)$ MHz.}
            \label{fig:sup:moddepth}
        \end{figure*}
        
Magnetic resonances are driven by frequency-modulated MW fields with an instantaneous frequency

\begin{equation}
    f(t)=f_\text{c}+f_\text{dev}\cos{2\pi f_\text{mod}t},
\end{equation}
where $f_\text{c}$ is the carrier frequency, $f_\text{dev}$ the modulation depth and $f_\text{mod}$ the modulation frequency. After optimizing the power levels, we fine-tune the modulation depth $f_{\text{dev}}$ to maximize the magnetic sensitivity. To this end, ODMR spectra are recorded for various modulation depths, and the contrast, linewidth and SNLS are extracted from fits. As shown in FIG. \ref{fig:sup:moddepth}, the SNLS exhibits a distinct optimum at a modulation depth of $4.5\,$MHz. Increasing $f_{\text{dev}}$ enhances the contrast but simultaneously broadens the resonance due to modulation broadening \cite{DENG2006254}. Fitting the linewidth dependence with the expression \cite{DENG2006254}:

\begin{equation}
    \Delta\nu=A\Delta\nu_{\text{intr}}\left\{\left(\frac{f_{\text{dev}}}{2\Delta\nu_{\text{intr}}}\right)^2+5-2\left[4+\left(\frac{f_{\text{dev}}}{2\Delta\nu_{\text{intr}}}\right)^2\right]^{1 / 2}\right\}^{1 / 2},
\end{equation}
where $\Delta\nu$ is the measured linewidth, $\Delta\nu_{\text{intr}}$ the intrinsic linewidth (without modulation broadening) and $A$ a scaling factor, yields an intrinsic linewidth of $\Delta\nu_{\text{intr}}=8.02(3)$ MHz. The corresponding Rabi frequency $\Omega_R$ can be estimated using \cite{Dumeige.2019, foot2005}:

\begin{equation}
    \Delta \nu_{\text{intr}}=\frac{1}{\pi T_2^*} \sqrt{1+\frac{W_{\text{MW}}}{W_{\text{sat}}}}=\frac{1}{\pi T_2^*}\sqrt{1+\frac{2\Omega_R^2}{1/(\pi T_2^*)^2}},
    \label{eq:sup:lw_intr}
\end{equation}
where $T_2^*$ is the spin-dephasing time, $W_{\text{MW}}$ the microwave pumping rate, and $W_{\text{sat}}$ the microwave saturation rate. Inverting this relation gives a Rabi frequency of $\Omega_R = 5.2(1)\,$MHz. The modulation frequency is chosen empirically to be $f_{\text{mod}} = 8\,$kHz, corresponding to a local minimum in the signal noise floor. The LIA filter time constant is set to $78\,$µs, resulting in a measurement bandwidth of $1\,$kHz.

        \subsection*{Rate equation model}
        \label{sup:rateq}

        The linearizations used in equation (2) in the main text are derived from a Taylor series expansion around the threshold carrier density $N=N_\mathrm{th}$:
        \begin{equation}
        \label{eq:lin}
        \begin{cases}
        \begin{aligned}
        &G(N) = gN_{\mathrm{tr}}\ln\left(\frac{N}{N_{\mathrm{tr}}}\right)\approx g\frac{N_\mathrm{tr}}{N_\mathrm{th}}\left(\kappa_1+N \right) \\
        \\
        &R(N)=AN+BN^2+CN^3\approx \tau_1 + \tau_2 (N-N_\mathrm{th})\\
        \\
        &r_{\mathrm{sp}}(N)=\frac{gN_{\mathrm{tr}}\ln\left(1+\frac{N^2}{N_{\mathrm{tr}}^2}\right)}{2}\approx g\frac{N_\mathrm{tr}N_\mathrm{th}}{N_\mathrm{tr}^2+N_\mathrm{th}^2}\left(\kappa_2+N \right)
        \end{aligned}
        \end{cases},
        \end{equation}
        where $g$ denotes the differential gain, $N_{\mathrm{tr}}$ the transparency carrier density, $\tau_N$ the effective carrier lifetime and $A$, $B$, $C$ are recombination coefficients.
        The linearization constants are given by:
        \begin{equation}
        \label{eq:lin_const}
        \begin{cases}
        \begin{aligned}
        &\kappa_1 =  N_\mathrm{th} \left[ 
        \ln\left(\frac{N_\mathrm{th}}{N_\mathrm{tr}}\right) -1 \right] \\
        \\
        &\kappa_2 =  N_\mathrm{th} \left[\frac{N_\mathrm{tr}^2+N_\mathrm{th}^2}{2N_\mathrm{th}^2} 
        \ln\left(1+\frac{N_\mathrm{th}^2}{N_\mathrm{tr}^2}\right) -1 \right]\\
        \\
        &\tau_1=AN_\mathrm{th}+BN_\mathrm{th}^2+CN_\mathrm{th}^3\\
        \\
        &\tau_2=A+2BN_\mathrm{th}+3CN_\mathrm{th}^2
        \end{aligned}
        \end{cases}.
        \end{equation}
        With these relations, the explicit analytical steady-state solution for the optical power in the cavity mode $P_\alpha$, as a function of the injected current $I$ is given by:
        \begin{widetext}
        \begin{multline}
        \label{eq:pi}
P_\alpha(I) = \frac{1}{2 \beta_{\mathrm{g}} \gamma \left(\alpha - \Gamma \kappa_{\mathrm{1}} \sigma \right)} \left[ \sqrt{
 \begin{aligned}
    \left[\beta_{\mathrm{g}} \beta_{\mathrm{sp}} \gamma (-\kappa_{\mathrm{1}} + \kappa_{\mathrm{2}}) + I \delta (\beta_{\mathrm{g}} - \Gamma \sigma) + \gamma \delta \left(-\beta_{\mathrm{g}} \tau_{\mathrm{1}} - \alpha \tau_{\mathrm{2}}
     + \beta_{\mathrm{g}} (N_{\mathrm{th}} + \kappa_{\mathrm{1}}) \tau_{\mathrm{2}} +
     \Gamma \sigma (\tau_{\mathrm{1}} - N_{\mathrm{th}} \tau_{\mathrm{2}}) \right) \right]^2 \\
     -4 \beta_{\mathrm{g}} \beta_{\mathrm{sp}} \gamma \delta \left(-\alpha + \Gamma \kappa_{\mathrm{1}} \sigma \right) \left[I - \gamma \tau_{\mathrm{1}} + \gamma (N_{\mathrm{th}} + \kappa_{\mathrm{2}}) \tau_{\mathrm{2}} \right]
 \end{aligned}
 } \right. \\
 + \left. I \beta_{\mathrm{g}} \delta - \beta_{\mathrm{g}} \beta_{\mathrm{sp}} \gamma \kappa_{\mathrm{1}} + \beta_{\mathrm{g}} \beta_{\mathrm{sp}} \gamma \kappa_{\mathrm{2}} - I \Gamma \delta \sigma - \beta_{\mathrm{g}} \gamma \delta \tau_{\mathrm{1}} + \gamma \Gamma \delta \sigma \tau_{\mathrm{1}} - \alpha \gamma \delta \tau_{\mathrm{2}} + N_{\mathrm{th}} \beta_{\mathrm{g}} \gamma \delta \tau_{\mathrm{2}} + \beta_{\mathrm{g}} \gamma \delta \kappa_{\mathrm{1}} \tau_{\mathrm{2}} - N_{\mathrm{th}} \gamma \Gamma \delta \sigma \tau_{\mathrm{2}} \right].
        \end{multline}
        \end{widetext}
        The associated parameters are defined as:
        \begin{equation}
        \label{eq:decl}
        \begin{cases}
        \begin{aligned}
        &\beta_\mathrm{g} = \Gamma g \frac{N_\mathrm{tr}}{N_\mathrm{th}} \\
        \\
        &\beta_\mathrm{sp} = \frac{Kv_{\mathrm{g}}\hbar\omega\Gamma g}{l} \frac{N_\mathrm{tr}N_\mathrm{th}}{N_\mathrm{tr}^2+N_\mathrm{th}^2} \\
        \\
        &\gamma =  edWl\\
        \\
        &\delta = \eta_\mathrm{i}dW \hbar\omega
        \end{aligned}
        \end{cases}.
        \end{equation}
        All simulation parameters are listed in TABLE \ref{tab:parameters}. 
        Using equation (\ref{eq:pi}) in equation (3) in the main text, we fit the experimentally obtained off-resonant power-current characteristic of the LICAM configuration to extract $\eta_\mathrm{i}$, $P_\mathrm{sp}$ and $N_\mathrm{th}$. The threshold carrier density $N_\mathrm{th}$ is related to the threshold current $I_\mathrm{th}$ through:
    
        \begin{equation}
            \label{eq:current}
            I_\mathrm{th}=\gamma R(N_\mathrm{th}).
        \end{equation}
        From $N_\mathrm{th}$, we determine the effective absorption constant $\alpha$ using the threshold condition:
        
        \begin{equation}
            \label{eq:nth}
            \Gamma G(N_\mathrm{th})-\Gamma\sigma N_\mathrm{th}-\alpha=0,
        \end{equation}
        yielding $\alpha=6.05\,\text{cm}^{-1}$. This loss corresponds to a passive round-trip loss of $1-\exp{(-2\alpha l})\approx0.7$ or an average storage time of $1.4$ round trips per photon. Consequently, passive multipass absorption is not a relevant contribution to the observed LICAM enhancement, which is instead governed by the active laser dynamics captured by the rate-equation model. The effective absorption constant $\alpha$ is composed of internal chip losses, mirror losses $-\frac{1}{2l}\ln (R_\mathrm{f}R_\mathrm{r})\approx5.11\,\text{cm}^{-1}$, external cavity losses (e.g. from the diamond and lens), and additional differential losses $\Delta\alpha$ induced by resonant IR absorption in the diamond. Note, that the rate equation model treats the laser system effectively as homogeneous system of length $l$. In the single-pass configuration, resonant differential absorption $\Delta\alpha$ increases the optical depth by $\tau_\mathrm{r}$ and reduces the output power from $P_\mathrm{off}$ to $P_\mathrm{on}$. For small optical depths ($\tau_\mathrm{r}<<1$), we approximate

        \begin{equation}
            C_\mathrm{sp} = \frac{\Delta P}{P_\mathrm{off}} = 1-\exp\lbrace-\tau_\mathrm{r}\rbrace  \overset{\tau_\mathrm{r}<<1}{\approx} \tau_\mathrm{r},
        \end{equation}
        and identify the single-pass contrast $C_\mathrm{sp}=3.8\times10^{-5}$ with $\tau_\mathrm{r}$.
        This yields a differential absorption constant of $\Delta\alpha=\tau_\mathrm{r}/l=0.038\,\text{m}^{-1}$.
        On resonance $(\alpha \rightarrow\alpha+\Delta\alpha)$, equation (\ref{eq:nth}) can be inverted for $N_\mathrm{th}^\mathrm{(on)}$:

        \begin{equation}
            N_\mathrm{th}^\mathrm{(on)} = \mathrm{Re}\left\lbrace-\frac{g}{\sigma}N_\mathrm{tr}W_0\left\lbrace -\frac{\sigma}{g}\exp\left\lbrace\frac{\alpha+\Delta\alpha}{gN_\mathrm{tr}\Gamma}\right\rbrace \right\rbrace\right\rbrace,
        \end{equation}
        where $W_0$ denotes the principal branch of the lambert $W$ function. Substituting $N_\mathrm{th}^\mathrm{(on)}$ into equations (\ref{eq:pi}) and (\ref{eq:decl}) allows the calculation of the on-resonant output power $P_{\alpha+\Delta\alpha}(I)$ as a function of injection current $I$. All simulation parameters are listed in TABLE \ref{tab:parameters}. We use superscripts to denote if the listed parameter values are directly measured in the experiment($\dag$), experimental design parameters or known from the chip fabrication ($\triangle$), target values ($\oplus$), literature-consistent fixed model inputs ($*$) or extracted from fits ($\propto$). We assume an ideal mode without excess noise and therefore fix the Petermann K-factor to $K=1$ \cite{Petermann.1979,Siegman.1989}. The recombination parameters $A,B,C$ and the free carrier absorption cross section $\sigma$ are not  independently determined in the experiment. Instead we fix them, based on experience, within a literature-consistent order-of-magnitude range \cite{Kastner.2019, Coldren.2012, Delaney.2009, Krishnamurthy.2011}.
        
       \begin{table}[h]
            \centering
            \begin{tabular}{|p{3cm}|c|c|c|c|}
            \hline
            \textbf{Parameter} & \textbf{Symbol} & \textbf{Exp.} & \textbf{Prosp.} & \textbf{Unit} \\
            \hline
            Wavelength & $\lambda$ & $1042^{\dag}\,$\cite{Acosta.2010_broadband} &  & nm \\
            ODMR linewidth & $\Delta\nu$ & $9.267^{\dag}$ & $0.5^{\oplus}\,$\cite{TayefehYounesi.2025} & MHz \\
            Differential absorption constant & $\Delta\alpha$ & $0.038^{\dag}$ & $20^{\oplus}\,$\cite{TayefehYounesi.2025} & m$^{-1}$ \\
            Thickness of active region & $d$ & $15^{\triangle}$ &  & nm \\
            Width of active region & $W$ & $5^{\triangle}$ &  & µm \\
            Length of active region & $l$ & $1^{\triangle}$ &  & mm \\
            Rear facet reflectivity & $R_\mathrm{r}$  & $0.4^{\triangle}$ & $1$ &  \\
            Front facet reflectivity & $R_\mathrm{f}$ & $0.9^{\triangle}$ & optimized &  \\
            Internal efficiency & $\eta_{\mathrm{i}}$ & $0.244^{\propto}$ & 0.9 &  \\
            Background spontaneous emission power & $P_{\mathrm{sp}}$ & $2.47^{\propto}$ & $0$ & µW \\
            Background cavity losses & $\alpha_0$ & $1^{\dag}$& $0.5^{\oplus}\,$\cite{Wenzel.2021,Veselov.2019} & $\text{cm}^{-1}$ \\
            Confinement factor & $\Gamma$ & $0.005^{\triangle}$ &  &  \\
            Recombination parameter & $A$ & $2.6 \times 10^{9\,*}$ &  & s$^{-1}$ \\
            Recombination parameter & $B$ & $5\times10^{-16\,*}$ &  & m$^3$s$^{-1}$ \\
            Recombination parameter & $C$ & $4 \times 10^{-42\,*}$ &  & m$^6$s$^{-1}$ \\
            Differential gain & $g$ & $1900 \times 10^{-22\,\triangle}$ & optimized & m$^2$ \\
            Transparency carrier density & $N_{\text{tr}}$ & $1.7 \times 10^{24\, \triangle}$ &  & m$^{-3}$ \\
            Threshold carrier density & $N_{\text{th}}$ & $2.49 \times 10^{24\,\propto}$  &  & m$^{-3}$ \\
            Group index & $n_{\mathrm{g}}$ & $3.6^{\triangle}$ &  &  \\
            Cross section of free carrier absorption & $\sigma$ & $10 \times 10^{-22\,*}$ &  & m$^2$ \\
            Petermann's K-factor & $K$ & $1$ &  &  \\
            \hline
            \end{tabular}
            \caption{List of experimental and prospective parameter values used in the simulations. $^{\dag}$measured, $^\triangle$design/fabrication parameter, $^{\oplus}$target value, $^*$fixed model input, $^\propto$fit}
            \label{tab:parameters}
        \end{table}

        \subsection*{Parameter optimization}
        \label{sup:opt}

        \begin{figure}[t!]
            \centering
            \includegraphics[width=.49\textwidth]{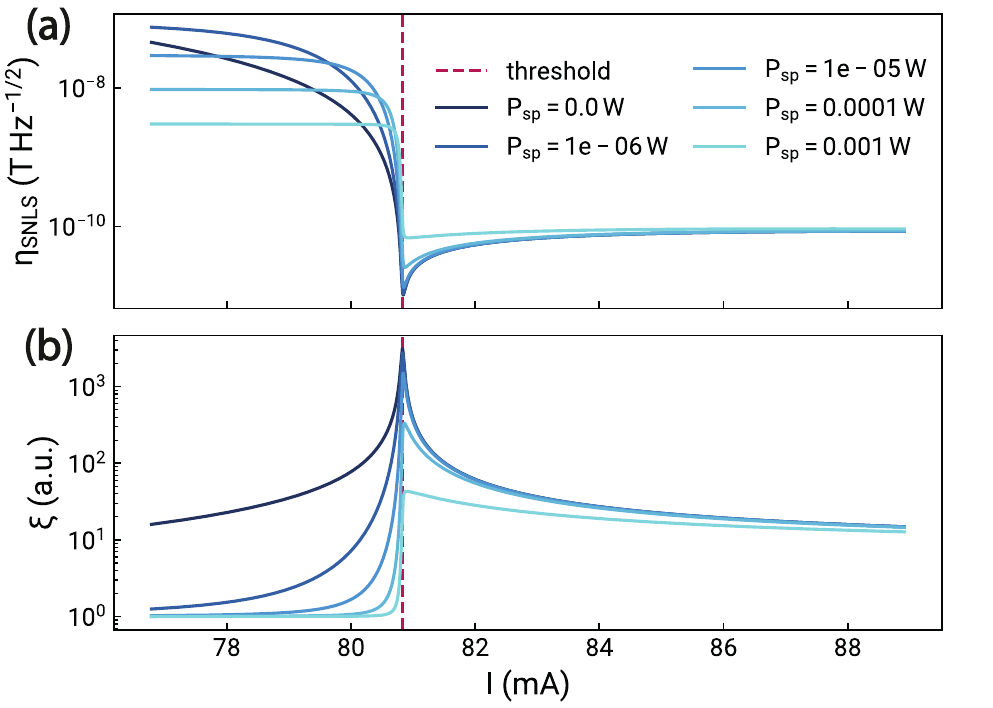}
            \caption[]{Simulated characteristic curves of (a) the SNLS and (b) the enhancement factor with respect to the gain current for different background spontaneous emission powers. The red dashed line marks the lasing threshold. Simulations are based on the improved primary parameters (TAB. \ref{tab:parameters}, prosp.) and $R_\mathrm{f}=0.5$.}
            \label{fig:sup:powers}
        \end{figure}

        The primary parameters we improve are the ODMR linewidth $\Delta\nu$, the background losses $\alpha_0$, the differential absorption constant $\Delta\alpha$, the background spontaneous emission power $P_\mathrm{sp}$ and the internal efficiency $\eta_\mathrm{i}$. Although the experimental enhancement factors exceed two orders of magnitude compared to the single-pass configuration, there remains significant potential for improvement. The main limitation of the current cross-configuration is the small overlap volume ($\approx300\,$µm in diameter) between the pump and probe beams. This results in a small optical depth, requiring high MW and optical pump powers to achieve a measurable absorption signal. However, these strong fields also induce power broadening of the ODMR resonances, yielding a linewidth of about $9.3\,$MHz in our setup. Extending the pump-probe region to span the entire IR propagation path through the diamond substantially increases the optical depth. This allows more efficient MW driving and the use of hyperfine-resolved resonances to reduce the linewidth. The green pump power requirement could be significantly relaxed by using a doubly resonant cavity geometry that improves the pump conversion efficiency \cite{Dumeige.2013}. Younesi et al. \cite{TayefehYounesi.2025} report single-pass IR absorption ODMR with a contrast of up to $1.6\%$ over $2.6\,\mathrm{mm}$ and linewidths down to approximately $700\,\mathrm{kHz}$, corresponding to $\Delta\alpha \approx 6.2\,\mathrm{m}^{-1}$. Motivated by this benchmark, we use $\Delta\alpha=20\,\mathrm{m}^{-1}$ and $\Delta\nu=500\,\mathrm{kHz}$ as realistic prospective target values in the simulations. With proper optical integration, the background losses can be further reduced from about $1\,\text{cm}^{-1}$ to $0.5\,\text{cm}^{-1}$\cite{Wenzel.2021,Veselov.2019}, and the internal efficiency can be increased to $\eta_\mathrm{i}=0.9$. Improvements in these primary parameters directly enhance both the SNLS and the enhancement factor. This effect is illustrated in FIG. \ref{fig:sup:powers}, which shows simulated characteristic curves of the SNLS and enhancement factor as a function of the gain current for different values of the background spontaneous emission power $P_\mathrm{sp}$. Since $P_\mathrm{sp}=2.47\,$µW in the measurements is negligible for the overall performance, it is set to $0$ in the simulations. Diamond material optimization provides an additional route to improve the primary parameters \(\Delta\alpha\) and \(\Delta\nu\). \(\Delta\alpha\) can be directly increased through higher NV\(^{-}\) density, improved N-to-NV\(^{-}\) conversion efficiency, and a larger NV\(^{-}\) charge-state fraction, provided that additional background absorption and spin-dephasing remain controlled. At the same time, \(\Delta\nu\) is reduced by suppressing inhomogeneous broadening from strain, residual substitutional nitrogen, \(^{13}\mathrm{C}\) nuclear spins, and vacancy-related paramagnetic defects. Recent high-sensitivity NV ensemble magnetometers have demonstrated that tailored diamond growth, high \(^{12}\mathrm{C}\) isotopic purity, low-strain material, thermal stabilization, and spin-bath control can substantially improve the relevant dephasing and contrast parameters \cite{Barry.2020,Barry.2024}. 

        \begin{figure}[t!]
            \centering
            \includegraphics[width=.49\textwidth]{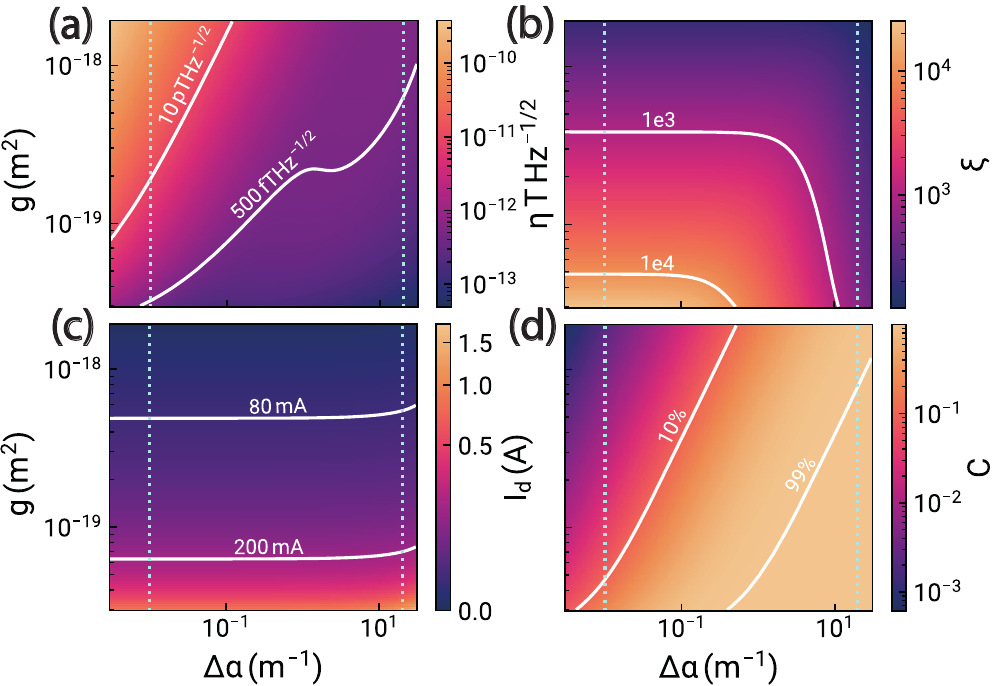}
            \caption[]{LICAM simulations of (a) the SNLS, (b) the enhancement factor, (c) the threshold current and (d) the spin contrast at the local optimum above the lasing threshold with respect to the differential gain $g$ and the differential absorption constant $\Delta\alpha$. Blue dashed lines indicate two configurations ($\Delta\alpha=20\,\text{m}^{-1}$ and $\Delta\alpha=0.01\,\text{m}^{-1}$) used for further optimization. Simulations are based on the improved primary parameters (TAB. \ref{tab:parameters}, prosp.) and $R_\mathrm{f}=0.5$.}
            \label{fig:sup:a_vs_g}
        \end{figure}

\begin{figure}[t!]
            \centering
            \includegraphics[width=.49\textwidth]{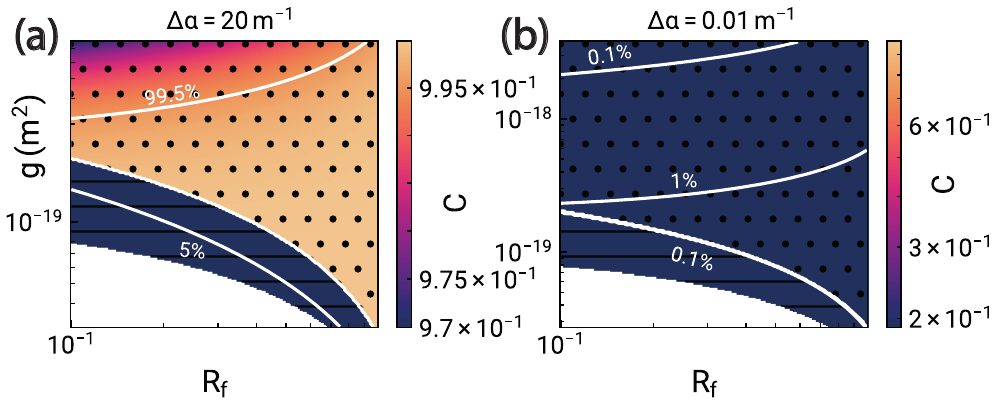}
            \caption[]{LICAM simulations of the spin contrast as a function of differential gain $g$ and front facet reflectivity $R_\mathrm{f}$ for on-resonant absorption constants of (a) $\Delta\alpha=20\,\text{m}^{-1}$ and (b) $\Delta\alpha=0.01\,\text{m}^{-1}$. The shading indicates whether the lasing threshold is reached (dot) or not (horizontal line).}
            \label{fig:sup:g_vs_R}
        \end{figure}

        The secondary parameters optimized are the mirror reflectivities $R_\mathrm{r}$, $R_\mathrm{f}$ and the differential gain $g$. Because light exiting the cavity through the rear facet does not contribute to detection, we set $R_\mathrm{r}=1$ and optimize only $R_\mathrm{f}$. Next, we vary $g$ and $\Delta\alpha$ while monitoring the SNLS, the enhancement factor, the spin contrast and the gain current at the local optimum around the lasing threshold. FIG. \ref{fig:sup:a_vs_g} displays the results of this simulation. The SNLS improves with increasing $\Delta\alpha$ and decreasing $g$, as the spin contrast approaches unity. However, these conditions require a larger threshold current. In the simulations we limit the gain current to $200\,$mA, corresponding to the experimental limit of the ECDL chip. Significantly larger enhancement factors are observed for smaller $\Delta\alpha$, where the spin contrast remains below saturation. We investigate two representative configurations: High differential absorption with $\Delta\alpha=20\,\text{m}^{-1}$, leading to an improved SNLS. Small differential absorption with $\Delta\alpha=0.01\,\text{m}^{-1}$, leading to a sharper enhancement compared to the single-pass configuration. For both configurations, we further optimize the SNLS and enhancement factor with respect to $g$ and $R_\mathrm{f}$. The results are presented in FIG. 5 in the main text, with the corresponding spin contrast shown in FIG. \ref{fig:sup:g_vs_R}. Three operational regimes are distinguished within the gain current limit: The system does not reach the lasing state within the accessible range, the optimum performance occurs at threshold, and the optimum performance occurs at the maximum gain current. We find optimum SNLS values of  $22\,\text{fT}\,\text{Hz}^{-1/2}$ for $\Delta\alpha=20\,\text{m}^{-1}$ and $2\,\text{pT}\,\text{Hz}^{-1/2}$ for $\Delta\alpha=0.01\,\text{m}^{-1}$.

        \subsection*{Operation as magnet sensor}

\begin{figure*}[!htpb]
            \centering
            \includegraphics[width=.99\textwidth]{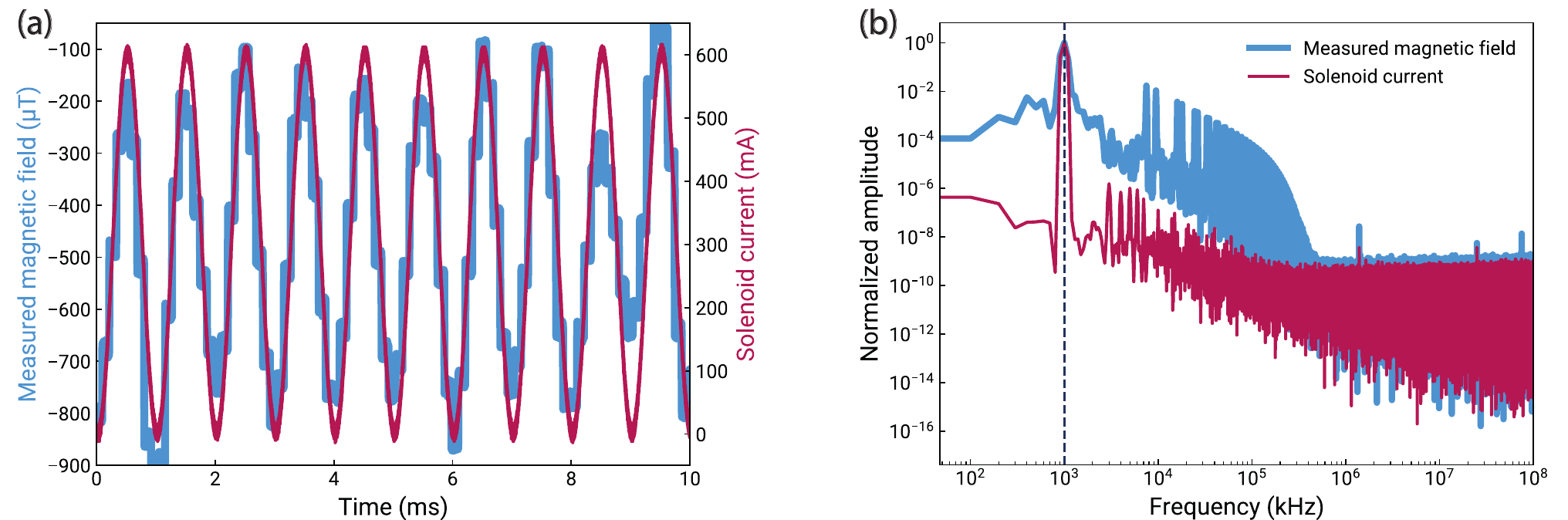}
            \caption{Demonstration of the LICAM system as a magnet field sensor. A $1\,$kHz sinusoidal current (red) applied to a copper coil generates an AC magnetic field, which is detected by the LICAM sensor (blue). (a) Time-domain segment of the measured and reference signals. (b) Corresponding frequency-domain representation showing the $1\,$kHz peak.}
            \label{fig:operation}
        \end{figure*}

        Finally, the LICAM system is demonstrated as a magnetic field sensor. A sinusoidal magnetic field at $1\,$kHz is generated by driving a copper coil positioned near the diamond. The sensor operates at a relative gain of $1.01$ and a fixed MW frequency of $2.7415\,$GHz with an input power of $32\,$dBm. The test signal is recorded with a $1\,$kHz bandwidth of the LIA, using frequency modulation at $8\,$kHz with a modulation depth of $4.5\,$MHz. FIG. \ref{fig:operation} shows the time-domain signal and the corresponding Fourier transform of the measured and reference signals. The $1\,$kHz peak is clearly visible in the measured spectrum, confirming that the LICAM sensor successfully detects the $1\,$kHz magnetic field generated by the solenoid.

        \subsection*{Noise mitigation strategies}

        As evident from FIG. 3(c) in the main text, the measured sensitivities near the lasing threshold remain over two orders of magnitude above the shot-noise limit. We attribute this discrepancy to technical noise sources that become especially relevant close to the threshold. In the following, possible noise mitigation strategies are discussed to mitigate these contributions.  
        
        Sources of noise include electronic noise in the signal detection chain, magnetic noise, laser intensity and polarization noise of the green laser and the ECDL-system, as well as mechanical vibrations of the external cavity.

        Electronic noise is expected below the shot-noise level and can therefore be neglected \cite{Bopp.2025, TayefehYounesi.2025}. Magnetic noise can arise from ambient magnetic-field fluctuations, fluctuations in the environmental spin bath of the NV ensemble, or mechanical vibrations changing the position of the permanent magnet with respect to the diamond. The latter can be improved by using mechanically decoupled Helmholtz coils to generate the bias field. More generally, decoupling from magnetic noise is possible by using pulsed sensing techniques \cite{Barry.2020}. However, comparing the on- and off-resonant noise floor in FIG. 3(b) in the main text shows that magnetic noise is not the dominant source in the experiment.

        Power and polarization fluctuations of the green pump laser translate into fluctuations of the ODMR contrast. In addition, pump-power fluctuations can induce thermal fluctuations in the diamond, that may affect optical alignment of the external cavity. A straightforward way to mitigate these mechanisms would be active power stabilization using an acousto-optic modulator in a feedback loop \cite{Tricot.2018} together with a polarization defining element.  

        Mechanical vibrations affect the cavity alignment by inducing cavity-length fluctuations and angular jitter of the output coupler. Both effects can translate into intensity fluctuations of the lasing mode. In addition, air currents can lead to refractive index fluctuations within the free-space cavity path \cite{Liu.2024}. We therefore expect that further miniaturization and integration of the ECDL-chip and external cavity into one mechanically rigid body would substantially improve passive stability. Active stabilization of the cavity using the IR output is generally challenging, as it is difficult to distinguish between power fluctuations and ODMR-induced absorption.  

        Gain and temperature fluctuations within the ECDL-chip also lead to intensity noise of the lasing mode. A powerful noise-rejection method for optical absorption measurements is balanced photodetection, where the signal of a probe beam is subtracted from the signal of a reference beam sharing the same optical source, thereby rejecting common-mode intensity noise \cite{Hobbs.1991}. Experiments employing this technique can realize operation near the shot-noise limit \cite{Acosta.2010_broadband, Bopp.2025, TayefehYounesi.2025}. However, balanced detection is not directly applicable to the present single-mode LICAM configuration, as the absorber is inside the laser cavity. Splitting this single mode over two detection channels would suppress both common-mode noise and ODMR absorption as well. 
        
        As a solution we propose multimode-LICAM, where similar to multimode intracavity absorption spectroscopy \cite{Fjodorow.2024, Baev.1999}, the laser system supports multiple modes, of which only a subset interacts with the NV absorber. After the cavity, the modes could be separated spectrally, spatially or by polarization, and a differential signal could be obtained from the absorption-sensitive and -insensitive modes. Such a scheme would be beneficial if broadband gain and cavity fluctuations are correlated between the modes, while ODMR-induced absorption differs for the two detection channels. Furthermore, the system could operate far above the lasing threshold in more stable lasing regimes \cite{Fjodorow.2024, Baev.1999}. Mode competition could further enhance the signal, as ODMR-induced absorption would be anti-correlated, but could also induce anti-correlated mode-partition noise. Noise correlation measurements are therefore crucial, when realizing such a system. One possible implementation could use several longitudinal modes spanning the NV$^-$ singlet absorption line, where the output modes are separated with a grating or etalon. 

        \onecolumngrid
